\documentclass[preprint,12pt]{aastex}

\newcommand{\etal}{{\em et al.}}            

\newcommand{\nh}{N$_H$}

\shorttitle{Cen A X-ray Jet}
\shortauthors{Kraft \etal}

\begin{document}

\title{{\em Chandra} Observations of the X-ray Jet in Centaurus A}
\author{R. P. Kraft, W. R. Forman, C. Jones, S. S. Murray}
\affil{Harvard/Smithsonian Center for Astrophysics, 60 Garden St., MS-31, Cambridge, MA 02138}
\author{M. J. Hardcastle, D. M. Worrall}
\affil{University of Bristol, Department of Physics, Tyndall Ave., Bristol BS8 ITL, UK}

\begin{abstract}
We present high angular resolution X-ray images and 
spectra from two {\em Chandra} ACIS-I observations
of the X-ray jet in the nearby radio galaxy Centaurus A.  
We find that the X-ray emission from
the jet is composed of a low surface brightness diffuse component
extending continuously from within at least 60 pc of
the active nucleus into the NE radio lobe 4 kpc from the nucleus, along with
31 discrete compact knots, most of which are extended
at the resolution of our observation.
We find that there are small but significant differences between the
X-ray and radio morphologies of the inner jet at the arcsecond level,
making it unlikely that a single, spatially coincident population of
ultrarelativistic electrons is responsible for the emission in both energy regimes.  
We suggest that the X-ray knots of the inner jet are indeed the sites of particle
acceleration and shocks, and the X-ray and radio knot offsets are caused
by a combination of particle diffusion and energy loss.  
These offsets may be a common feature of all jets in
radio galaxies, or at least jets in FR I type galaxies, and may be
fundamental to the physics of such jets.
They are best observed in Cen A because the source is so close.
Even though the X-ray and radio knots are offset in position and there
are variations of more than a factor of three in the ratio of X-ray to
radio flux density in the inner jet, the
radio to X-ray two-point spectral indices at the X-ray knots are
not unusually flat, and are consistent with those
observed in other X-ray jets seen in FR I galaxies such as M87 and 3C 66B.
We find the width of the jet in the X-ray bandpass
to be narrower to that measured in the radio along its most of its length. 
The X-ray spectra of several regions of the jet are well fit by absorbed
power-law models with photon indices $\sim$2.2-2.5, although the spectrum
of one bright knot located $\sim$1 kpc from the nucleus (knot B)
is harder (photon index = 2.0).
\end{abstract}

\keywords{galaxies: active - galaxies: individual (Centaurus A, NGC 5128) 
galaxies: jets - X-rays: galaxies}

\section{Introduction}

Centaurus A (NGC 5128) is the nearest (d=3.4 Mpc) active galaxy to
the Milky Way and has been well studied across the entire
electromagnetic spectrum \citep{isr98}.
Its proximity allows the features of an active galaxy to
be studied at the highest linear spatial resolution ($1''\sim$17 pc
at the distance of Cen A).  Optically Cen A appears to be an elliptical galaxy crossed
by a dark dust lane through the center,
which is thought to be the result of a merger
with a small spiral galaxy \citep{sch94}.  Previous radio observations
of Centaurus A have detected a bright radio core, a superluminal nuclear
jet \citep{tin98}, a one-sided, kiloparsec scale jet extending $6'$
NE of the nucleus \citep{bur83}, two radio lobes
NE and SW of the nucleus, and extended, low
surface brightness diffuse emission spanning several degrees
on the sky.  It is often considered to be the proto-typical 
Fanaroff-Riley class I low-luminosity radio galaxy.  

Prior to the launch of {\em Chandra}, X-ray emission had been detected from
the knots in the jets, or hotspots in the lobes, of seven
radio galaxies \citep{har01}, but we are far from a detailed
understanding of the mechanisms responsible for this emission.
This has partly been due to the limited angular resolution of
X-ray observatories preventing a detailed comparison of
jet features across a wide sample of galaxies.
This has changed with the launch of the {\em Chandra}
X-ray Observatory.  Recent {\em Chandra} observations of FR I radio galaxies
have shown that jet X-ray emission may be
a common feature of all such galaxies \citep{wor01,hrd01},
implying that the production of X-ray emission
may be fundamental to the nature of such jets.

Cen A is one of two FR I galaxies in which an X-ray jet
had been detected prior to {\em Chandra} (M87 being the other),
and, as the closest such galaxy, provides an opportunity to
study jet physics at the highest linear resolution.
Previous X-ray observations of Cen A have detected X-ray
emission along much of the length of the jet, and bright
enhancements at the locations of the radio knots of the
inner jet \citep{sch79,fei81,dob96,tur97}.
Based largely on the similarities of the X-ray
and radio morphologies, it has been argued that a single
population of ultra-relativistic electrons is responsible
for the emission in both the radio and X-ray regimes \citep{fei81}.
Recent {\em Chandra} HRC observations of Cen A suggested small
but significant differences in the spatial morphology of the jet
in the X-ray and radio bandpasses \citep{kra00}.

No optical or infrared emission from the jet at the positions
of the bright X-ray knots of the inner jet has been
seen.  The upper limits placed by HST WFPC2 and
NICMOS observations, however, are well above an extrapolation of the radio 
to X-ray flux \citep{mar00}.
The inner 1 kpc of the jet passes through the dust lane of
the galaxy, making detection of any optical component difficult.
Several optical filaments have been detected in
the vicinity of the jet and NE radio lobe, but their relationship
to the X-ray/radio jet, if any, is unclear \citep{bro85, mor91, mor92, isr98}. 
Infrared (1.25 $\mu$m) emission extending $\sim 10''$ from the nucleus along the position
angle of the jet has been detected, but it was argued that
while this emission is probably related to the jet, it is not
due to synchrotron radiation from relativistic particles \citep{joy91}.

In this paper, we present preliminary results from the analysis
of two $\sim$36 ks {\em Chandra} ACIS-I observations of
the jet in Cen A.  The unprecedented angular resolution
and sensitivity of {\em Chandra} have allowed us to resolve many features
of the jet not previously seen.  We present spectra of the brightest knots and diffuse
emission in the jet, and make a detailed comparison of the jet's X-ray and radio morphology.
The goal of this paper is to present images, spectra, and
preliminary results from these observations.  
A more detailed interpretation and discussion of these
results will be presented in a future publication.
Analysis of other aspects of these observations, including the
X-ray point source population, emission from the hot
interstellar medium, and emission from the radio lobes
also will be presented in separate publications \citep{kra01}.
We assume a distance of 3.4 Mpc to Cen A throughout this paper.

\section{Instrumentation and Observations}

The {\em Chandra} telescope has an imaging resolution
of better than $1''$ on-axis.  The 50\% encircled
energy diameter has been measured to be $\sim 0.6''$ \citep{pog03},
and sources separated by $\leq 0.5''$ on-axis have
been resolved \citep{cha00}.  The imaging degrades significantly for 
sources more than $1'$ off-axis.
At $3'$ from the center of the field of view, 90\% of the
encircled energy is contained within a diameter of $3''$.
The sensitivity of the ACIS-I instrument extends from 0.4 keV
to 10 keV.  A more detailed discussion of the capabilities of the
{\em Chandra}/ACIS-I instrument is contained elsewhere \citep{wei00,pog03}.
The pointing and SIM position for each observation were chosen so that the
central region of the galaxy, including the nucleus, the jet, 
the SW radio lobe, and most of the point sources and diffuse
emission from the hot ISM were located within one CCD (the I3 chip).

The raw events were filtered to include only ASCA grades 0,2,3,4, and
6.  All events below 0.4 keV and above 5 keV were removed.  The response
of the ACIS-I drops rapidly below 0.4 keV, so that most of the
events below this are either background or higher energy events
in the tail of the energy redistribution function.  Above 5 keV,
most of the events are either particle background events or
events in the PSF wings from the bright nucleus.
All events at node boundaries were removed because of uncertainties
in grade reconstruction.  All events with pulse height invariant
channel (PI) equal to 0, 1, or 1024 were removed as they represent
unphysical signals; hot CCD columns and pixels were removed using the
standard table provided by the Chandra X-ray Center (CXC).  Short
term transients due to cosmic rays which produced
events in three or more consecutive frames
that could mimic a point source (van Speybroeck 2000, private
communication) also were removed.

\begin{figure}
\plotone{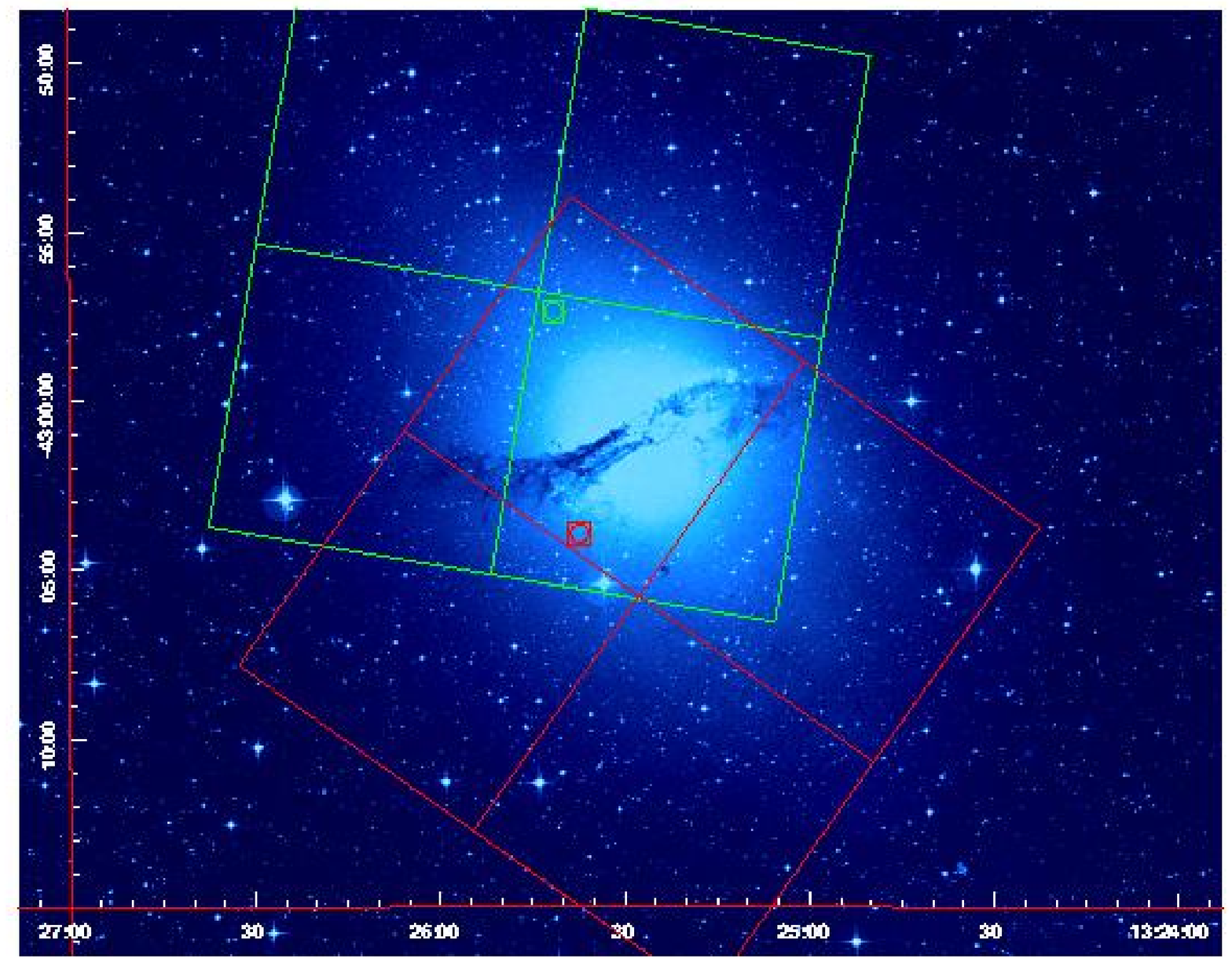}
\caption{The FOV of the two {\em Chandra} observations (OBSID 00316 - white, OBSID 009
62 - red) overlaid onto an optical DSS image (J band) of Cen A.  Both observations were
aligned so that the central region of Cen A, including all of
the X-ray jet, was well centered on the ACIS I3 CCD.  Coordinates are J2000.}\label{fov}
\end{figure}

Cen A was observed twice, on December 5, 1999 and May 17, 2000, with
the ACIS-I instrument.  The exposure times for the two observations (OBSIDs 00316 and 00962)
were 35856 s and 36510 s.  The data were examined for
periods of high background or problems with the
aspect solution, but none were found.
The aspect solution for {\em Chandra} observations is generally
good to $2''$ or better \citep{ald00}, but  
we have independently verified the aspect solution of both data sets
by comparing the positions of X-ray point sources at the
edge of the field of view with stellar positions tabulated in
the USNO A2.0 catalog \citep{mon98}, and estimate that our
absolute astrometry is accurate to within $\sim0.5''$ \citep{kra01}.
The FOV for each of the observations and the position
of best focus is shown in Figure~\ref{fov} superimposed on an
optical DSS image.

\begin{figure}
\plotone{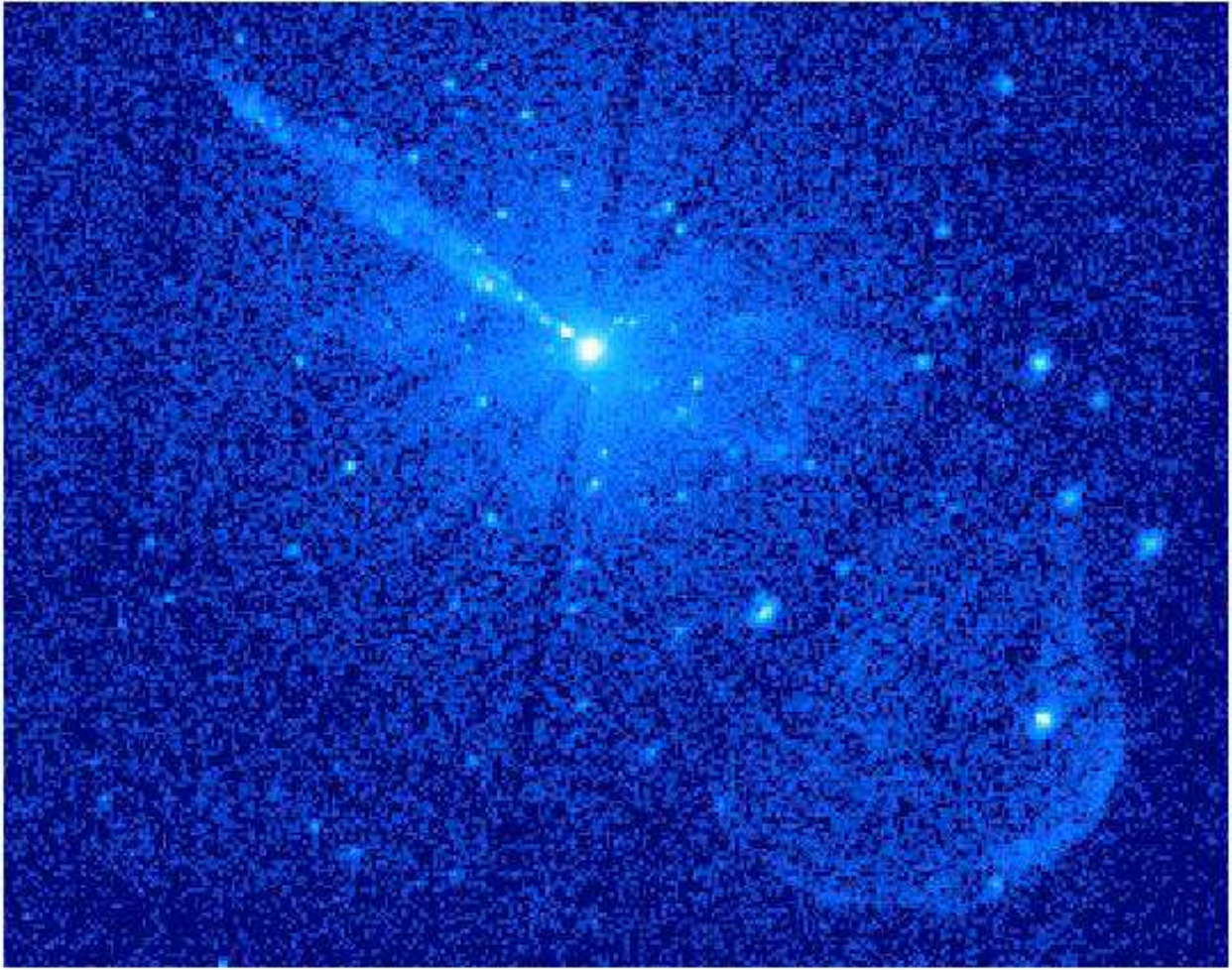}
\caption{Raw X-ray image of Centaurus A in 0.4-5 keV bandpass binned at
$2''$ per pixel.  The data from both observations have been co-added.  
The stripes radiating outward from the nucleus are the result of
removing the frame-transfer streak in each observation due to
out of time events from the nucleus.  North is up and east is to
the left.  The image scale is approximately $10'\times8'$.}\label{xraw}  
\end{figure}

\begin{figure}
\plotone{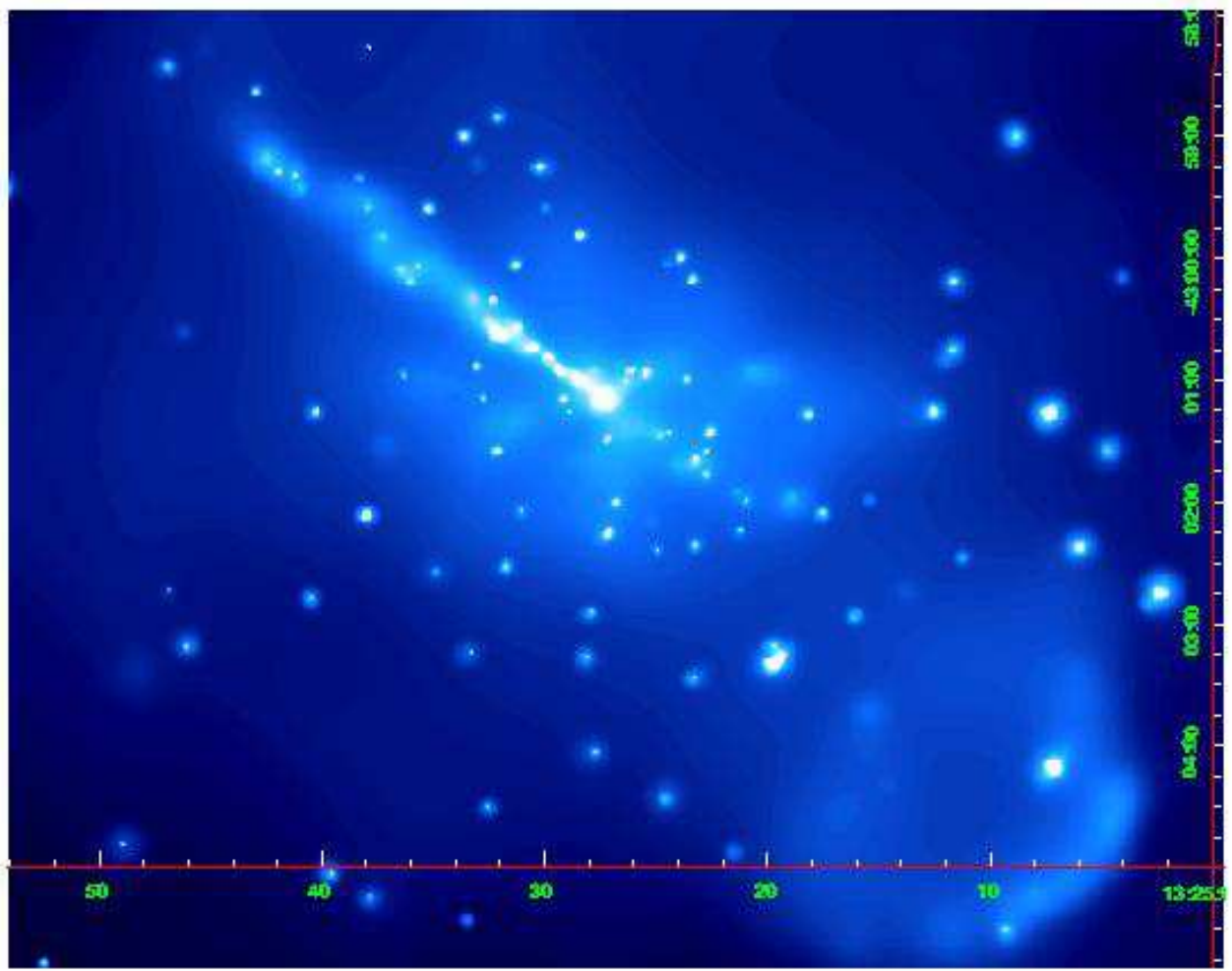}
\caption{Adaptively smoothed, co-added, exposure-corrected X-ray image 
of Centaurus A in the 1-3 keV band.  The nucleus is
the bright source near the center of the field, and the jet
extends to the NE.  North is up and east is left.  Note the
large number of point sources associated with the galaxy, the
extended diffuse emission from the hot ISM, and emission from
the SW radio lobe.}\label{cena}  
\end{figure}

The raw, co-added X-ray image in the 0.4-5 keV bandpass binned at $2''$
per pixel is shown in Figure~\ref{xraw}.
The nucleus is clearly visible, along with
the X-ray jet extending to the NE, many point sources,
emission from the hot ISM, and emission coincident with
the SW radio lobe.  This image has not been exposure corrected.
The radial stripes pointing from the
nucleus are the result of removing the frame transfer streak
caused by out of time events.  The other linear features around
the image are gaps between the various CCDs in one or the other of the
exposures.  An adaptively smoothed, co-added, exposure-corrected 
X-ray image of Cen A in the 1-3 keV band is shown in Figure~\ref{cena}.

We also use radio observations of Cen A made at 8.4 GHz between October
1990 and November 1991 with the NRAO Very Large Array
(VLA)\footnote{The National Radio Astronomy Observatory is a facility
of the National Science Foundation operated under cooperative
agreement by Associated Universities, Inc.}, consisting of
approximately two hours on-source integration time in each of the A,
B, C and DnC arrays. The data were reduced in the standard manner
using AIPS and then combined (after correcting for variations in the
flux density of the core) to produce a single $uv$ dataset, sensitive
to structure with a largest scale of $\sim 100''$ and with a
maximal resolution of $0.9'' \times 0.2''$. The radio maps
presented in this paper are all made from this full dataset with
appropriate tapering and weighting of the $uv$ plane. The nominal
dynamic range of the maps (peak to off-source r.m.s. noise) is about
10$^4$, but their fidelity is limited by residual phase and amplitude
artifacts about the 7 Jy core.  
We compared the position of the radio core in this data set with its
position in archival Australia Telescope Compact Array (ATCA)
observations at a similar frequency, and found that there was a small
($\sim 2''$) offset between the two data sets. We attribute this
difference to the use of a distant ($\sim 30$ degree) phase calibrator
in the VLA observations, together with the heavy use of
self-calibration in the data reduction; the ATCA observations, which
use a nearby phase calibrator and were only lightly self-calibrated,
are likely to have better positional accuracy. We have therefore
shifted the VLA data so that the core position agrees with that
determined from the ATCA data. After this adjustment, we find that the
X-ray and radio positions of the nuclei and the first bright
X-ray/radio knot AX1/A1 agree to within the absolute aspect
uncertainty of $\sim 0.5''$ of the X-ray data.

\section{Results}

\subsection{X-ray Morphology of the Jet}

\begin{figure}
\plotone{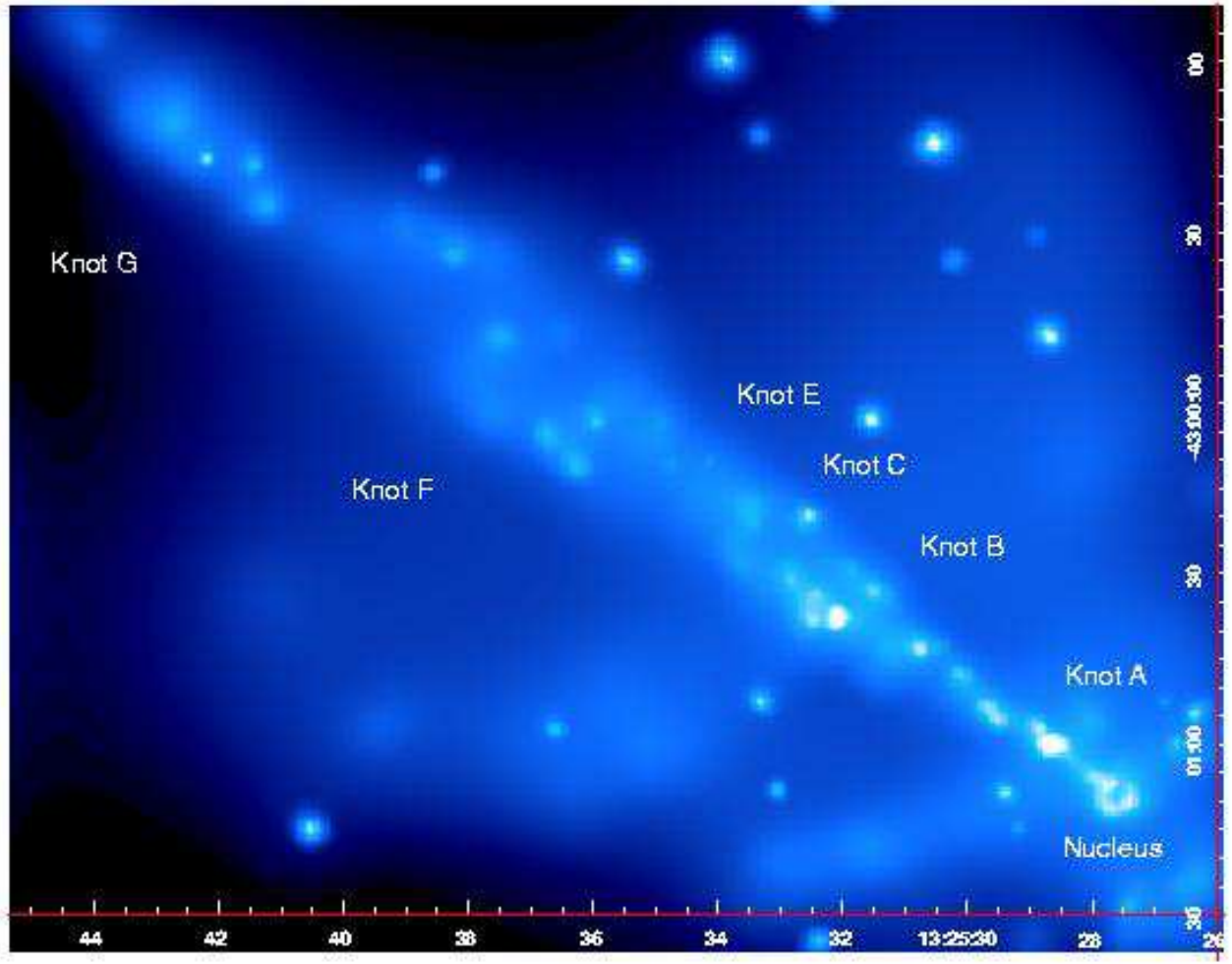}
\caption{Adaptively smoothed, co-added image
in the 0.4-2.5 keV bandpass of the Centaurus A jet.  North is up and east is
to the left.  The unusual donut shape of the nucleus to the southwest
is due to pile-up in the ACIS detector.  Each of the knots previously detected by
{\em Einstein} and ROSAT has been clearly resolved into several distinct subknots
and enhancements embedded within diffuse extended emission.  The knots and
enhancements have been labeled in the same manner as Table~\ref{knottab}.}\label{jetbw}
\end{figure}

\begin{figure}
\plotone{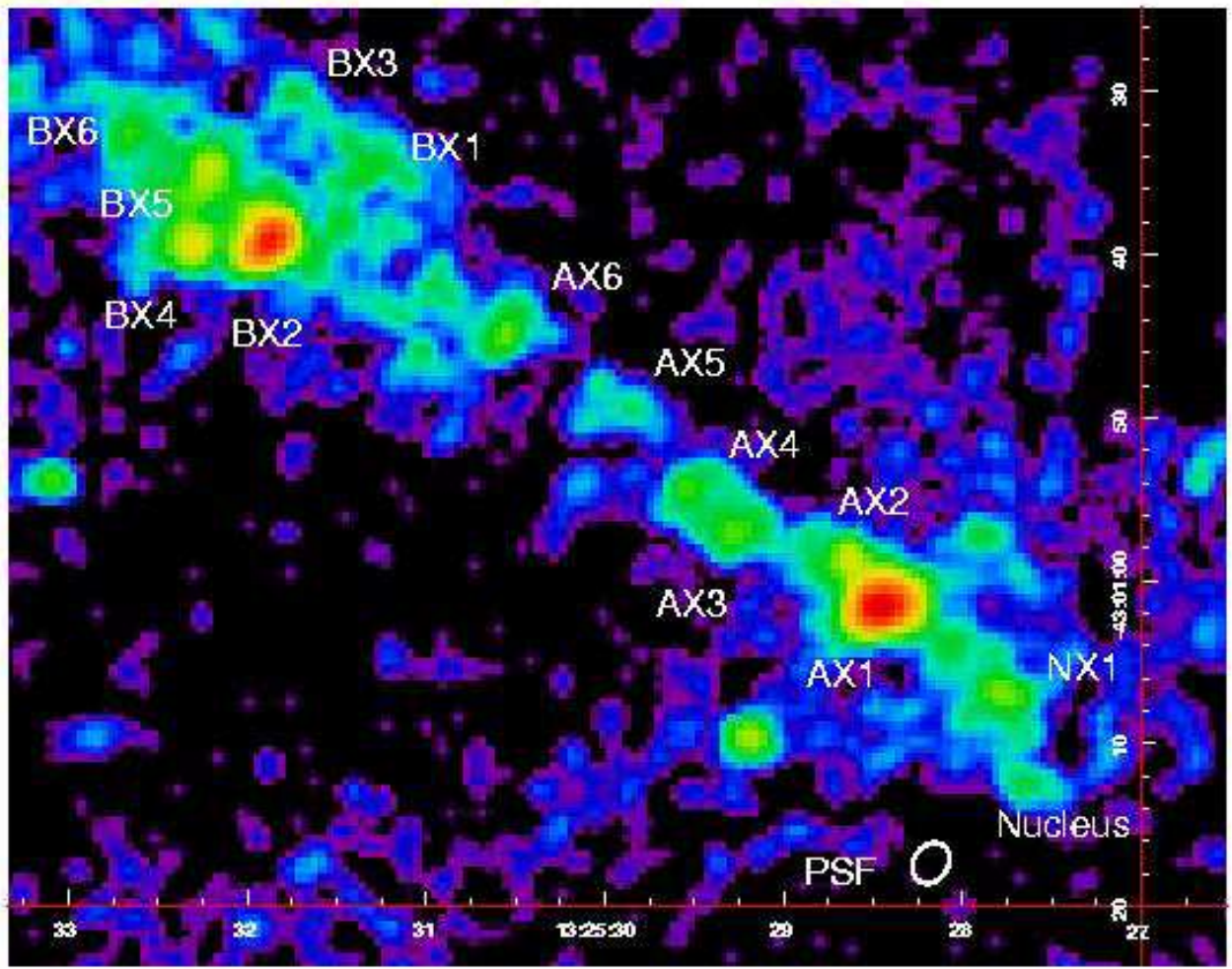}
\caption{Image of inner jet and radio knot B region of Centaurus A in the 0.4-1.5
keV bandpass.  This image contains data from only one (OBSID 00316) of the
observations and is in a softer band than Figure~\ref{jetbw}
to suppress the heavily absorbed nucleus.  The data have been smoothed by a $0.5''$ (r.m.s.)
Gaussian.  The X-ray knots have been labeled in the
same manner as in Table~\ref{knottab}.
The PSF (FWHM - smoothed by a $0.5''$ (r.m.s. Gaussian) to
match the data) at 1.5 keV at the position
of the nucleus based on MARX simulations (see text for full
discussion) is shown below the nucleus.}\label{gsjet}
\end{figure}

An adaptively smoothed, coadded, exposure corrected image of the Cen A jet in the 0.4
to 2.5 keV bandpass is shown in Figure~\ref{jetbw}.  
The nucleus is located in the SW corner of the image, and the
jet extends to the NE.  The count rate from the active nucleus
in the {\em Chandra} bandpass is
large ($\sim$6 cts s$^{-1}$) and the unusual shape of the nucleus in
this image is an artifact of pile-up in the ACIS-I detector.
Pile-up occurs when more than one event is incident on the
detector in the same or adjacent pixels in less than the frame time.
For moderate count rates (0.01 to 0.1 cts/s), this distorts the observed pulse height
spectrum, as there is no way to determine if a pixel contains multiple events.  At
high count rates, like that for the Cen A nucleus,
the event morphologies tend to migrate to grades that are not
telemetered to the ground and these are lost. The figure shows that diffuse emission
from the jet is observed continuously from the nucleus to the end of the 
X-ray jet $\geq 4'$ from the nucleus.
A slightly different view from the nucleus to knot B is shown in Figure~\ref{gsjet},
where the data in the 0.5 to 1.5 keV bandpass have been smoothed with a Gaussian ($\sigma$=$0.5''$).

\begin{figure}
\plotone{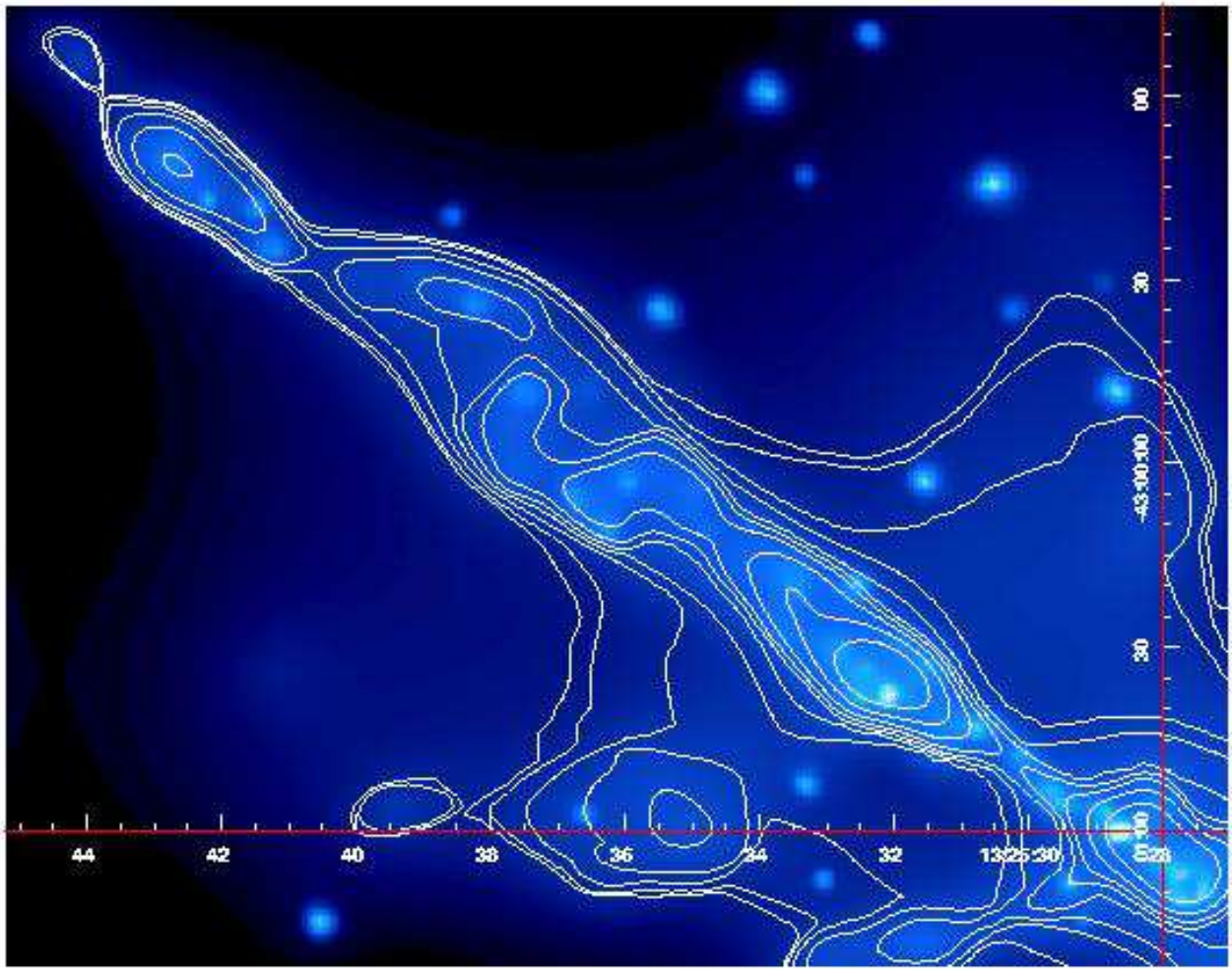}
\caption{Contour plot of the unresolved, diffuse emission of the X-ray jet
in the 0.4-2.5 keV band overplotted onto the adaptively smoothed image in
the same band (Figure~\ref{jetbw}).  The contours were created using a wavelet
decomposition, retaining only emission on scales of $8''$ or larger.  The
contours correspond to a surface brightness of 0.45, 0.51, 0.70, 1.02, 1.46, 2.03, 2.72, 3.55, 4.50, and 5.58$\times$10$^{-5}$ cts arcsec$^{-2}$ s$^{-1}$ above background.}\label{diffuse}
\end{figure}

Our {\em Chandra} observations show that the X-ray jet consists
of many knots of enhanced emission embedded within diffuse emission.
The diffuse emission extends continuously from the nuclear jet (knot NX1) into
the NE radio lobe beyond knot G.  Each of the previously 
detected X-ray knots has considerable substructure when viewed 
at higher resolution.  Using a wavelet decomposition source-detection
algorithm to detect emission enhancements
on angular scales of $0.5'', 1'', 2'',$ and $4''$ \citep{vik95}, we find
a total of 31 distinct knots or enhancements of X-ray emission embedded
within the diffuse emission.  
Previous {\em Einstein} observations \citep{fei81}
detected 7 distinct X-ray knots (labeled A to G), 
but subsequent ROSAT observations \citep{dob96}
were unable to confirm knot D.  Radio observations \citep{bur83}
showed that the inner knot (A) is actually composed of four subknots (A1 through A4).
We refer to the inner $45''$ of the jet, which includes these compact
radio knots, as the inner jet.
Figure~\ref{diffuse} contains a contour
map of the unresolved, diffuse emission from the jet with the
contribution from the knots removed overplotted onto the
adaptively smoothed image of the jet (Figure~\ref{jetbw}).
This contour map was created using a wavelet decomposition and
retaining only emission on scales of $8''$ and larger.

As described in detail below, many of the radio and X-ray knots of the inner
jet region do not exactly coincide, and all of the previously
reported knots beyond the inner jet are actually composed of
several smaller subknots.  To avoid confusion,
we will refer to the X-ray knots of the inner jet
as AX1, AX2, etc., and knots beyond the inner
jet as BX1, BX2, CX1, CX2, etc., depending on their location
relative to the knots of \citet{fei81} to clearly distinguish them from the
radio knots.  The nomenclature of \citet{bur83} will be used when referring specifically
to the radio knots.  A summary of previously known
radio and X-ray knots and their distances from the
nucleus is contained in Table~\ref{rxknots}.
When discussing X-ray jets, it is common to use the term
{\it knot} to specifically describe the sites of enhanced emission where
shocks may be accelerating the emitting particles.  We have detected
31 sites of enhanced emission,
but it is certainly not clear that they are all
sites of shocks and particle acceleration, particularly the
enhancements beyond knot B.
For simplicity, however, we will use the term {\it knot} throughout this
paper to describe these enhancements.

\begin{figure}
\plotone{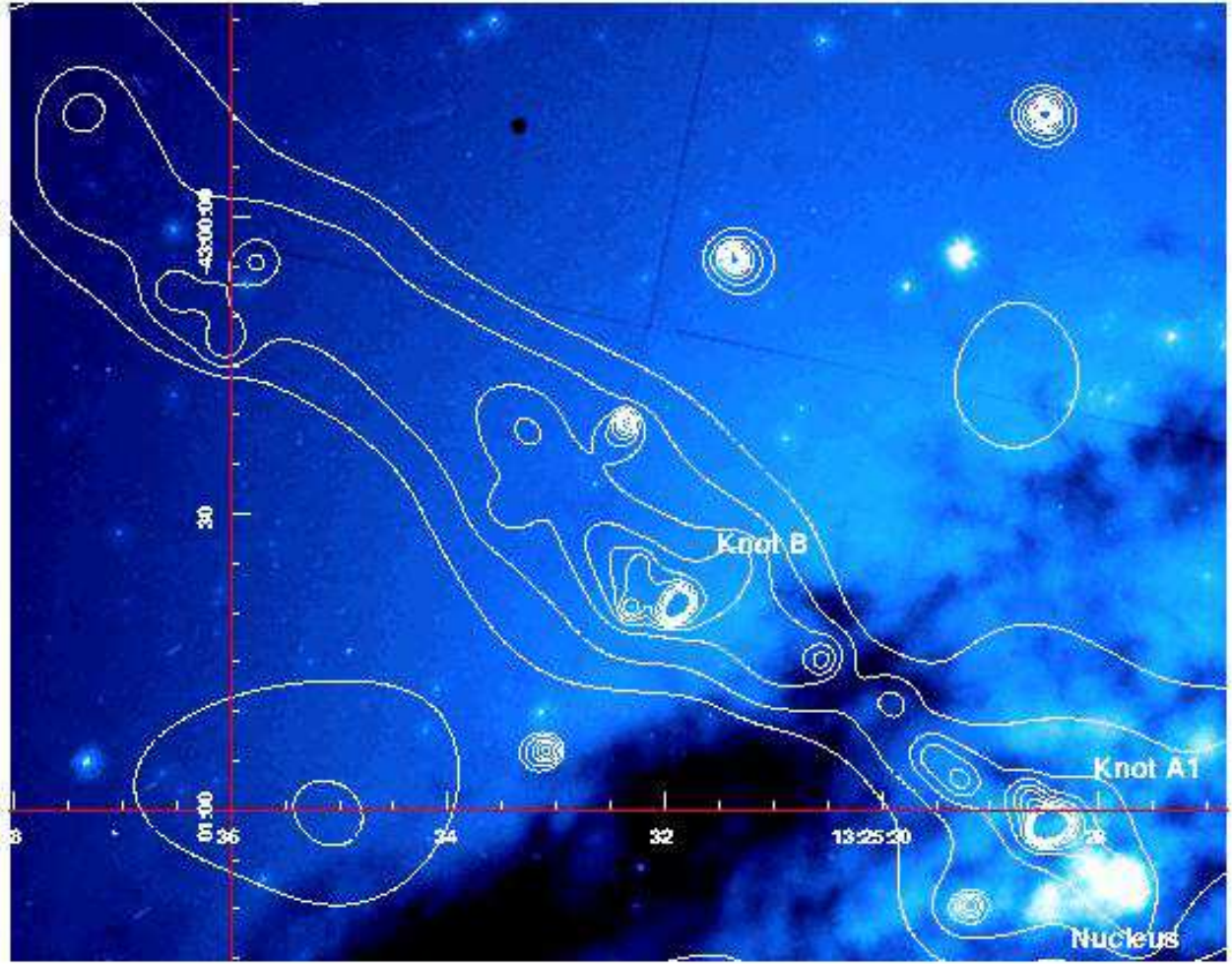}
\caption{Adaptively smoothed X-ray contours of the jet
in the 0.5 to 1.5 keV bandpass overlaid on a 350 s HST/WFPC (F675W filter -
central wavelength 6731 \AA) image
taken from the HST archive.  The X-ray contours correspond
to a surface brightness of 9.22, 16.3, 37.1, 72.2, 121, 184,
261, 352, 457, and 576$\times$10$^{-6}$ cts s$^{-1}$ arcsec$^{-2}$.
The background rate at a region adjacent to the end of
the jet (i.e. beyond any significant contribution from the hot ISM of the galaxy)
is 2.07$\times$10$^{-6}$ cts s$^{-1}$ arcsec$^{-2}$.}\label{optovl}
\end{figure}

In the inner jet region, X-ray emission from the nuclear jet
(radio knot N1) is clearly detected within $\sim 3''$ of the
nucleus, and detectable X-ray emission is clearly present between the
nuclear jet and the first bright radio/X-ray knot A1/AX1 as
seen from Figure~\ref{gsjet}.
The position of the first bright X-ray knot AX1 coincides
(within our absolute uncertainty in astrometry $\sim 0.5''$)
with radio knot A1 (see below).  Knot AX1 is clearly
extended both along the jet and perpendicular to it.
For reference, the PSF of the observation at the position of the nucleus (FWHM)
is shown on the righthand side of Figure~\ref{gsjet}.
There is continuous diffuse emission and several X-ray
knots between radio knots A1 and B, and there is a
kink or bend in the jet direction between knots AX2 and AX3.  
An image of the X-ray contours in the 0.4-1.5 keV bandpass
overlaid onto an optical (HST) image of
Cen A is shown in Figure~\ref{optovl}.
The region of low X-ray surface brightness between
knots AX5 and AX6 corresponds with dark dust bands in the optical.
The jet appears to widen dramatically before knot B just as it
emerges from behind the dust lane, although it appears
to remain collimated.  Beyond knot B, the jet is composed of both knots and diffuse
emission.  The width of the jet is approximately constant out to
knot E ($\sim 2.5'$ from the nucleus), where the jet apparently
begins to narrow (see Figure~\ref{cena}), although whether this is a
real feature or just due to a decrease in the X-ray surface brightness is not clear.

Variable absorption in the inner jet region might
have an effect on the observed X-ray structure.  The optical extinction, $A_V$, in the
inner jet region varies between 0.5 and 3.0 magnitudes \citep{sch96}.
Adopting a standard Galactic gas-to-dust ratio, this
corresponds to an X-ray column of 0.9-5.6$\times$10$^{21}$cm$^{-2}$ \citep{smi98}.
For a power-law spectrum with photon index = 2.2, this variation in $N_H$ would
correspond to a variation of a factor of $\sim$2 in count rate in ACIS-I in the
0.4-2.5 keV bandpass of the ACIS-I, although in the
region from the core to AX4 (before the dark dust lanes are
encountered) the positional variations in X-ray count rate
contributed by the absorption are expected to be less than 10\%.
The actual relationship between the X-ray emission from the jet and any
intervening absorbing material is probably more complex for two reasons.  First,
the angle of the jet to the line of sight is poorly constrained,
so the geometric relationship of the jet relative to the gas associated with the 
optically obscuring dust is by no means clear.  
Second, the X-ray spectral
fitting (described below) suggests that the absorbing column of the
brightest knots of the inner jet (AX1-AX4) is considerably larger
($N_H\sim$6$\times$10$^{21}$cm$^{-2}$)
than one would expect based on the extinction map of
\citet{sch96}.  In fact, this X-ray column density is at the high end of the
range inferred from $A_V$ over most or all of the inner jet.
We note that the morphological features of the inner jet
seen in the broad-band X-ray image are still present when
X-rays in the 1-3 keV band only are selected, so the appearance
of the inner jet is probably not significantly distorted because of variable
absorption.

\begin{figure}
\plottwo{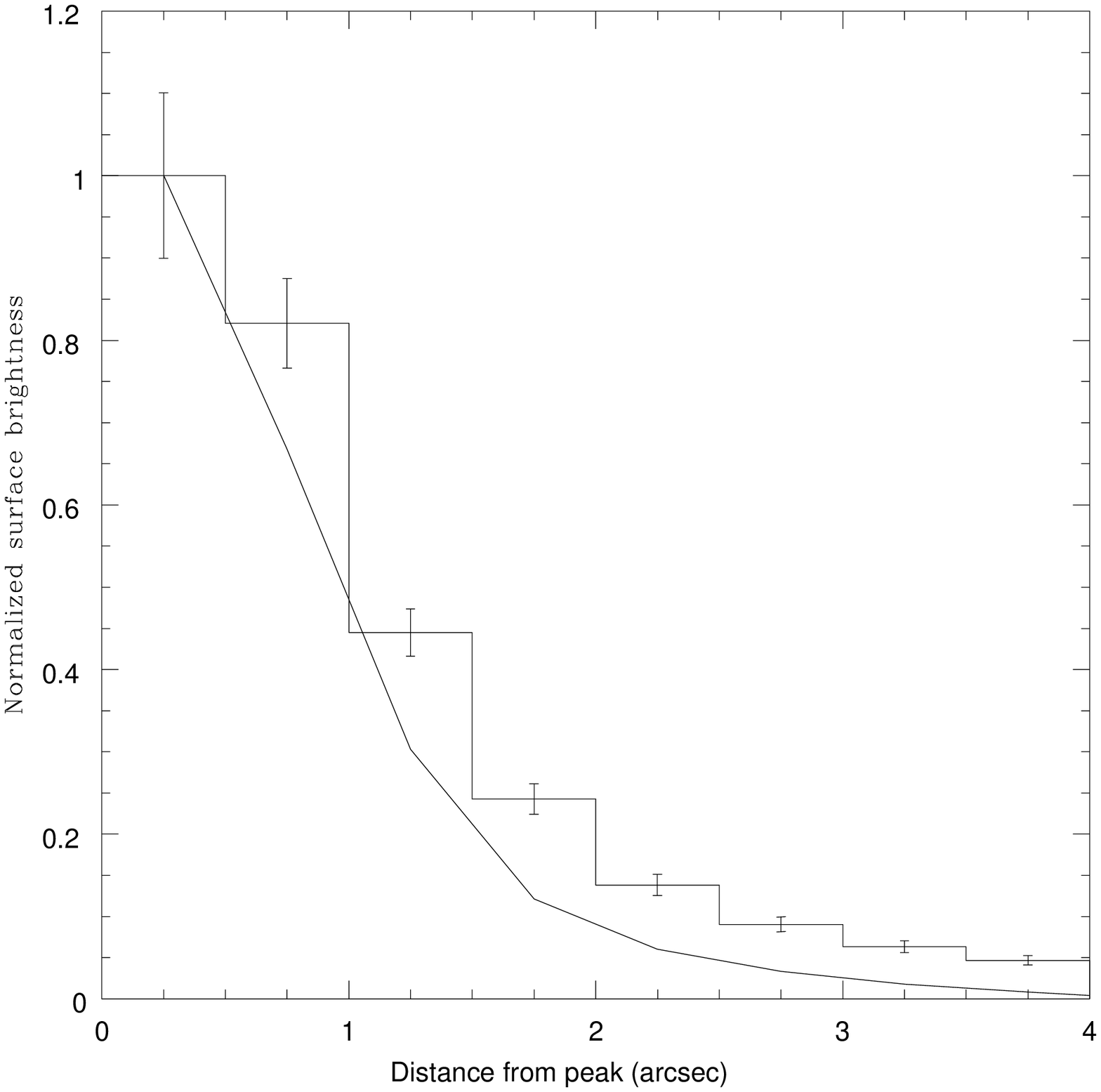}{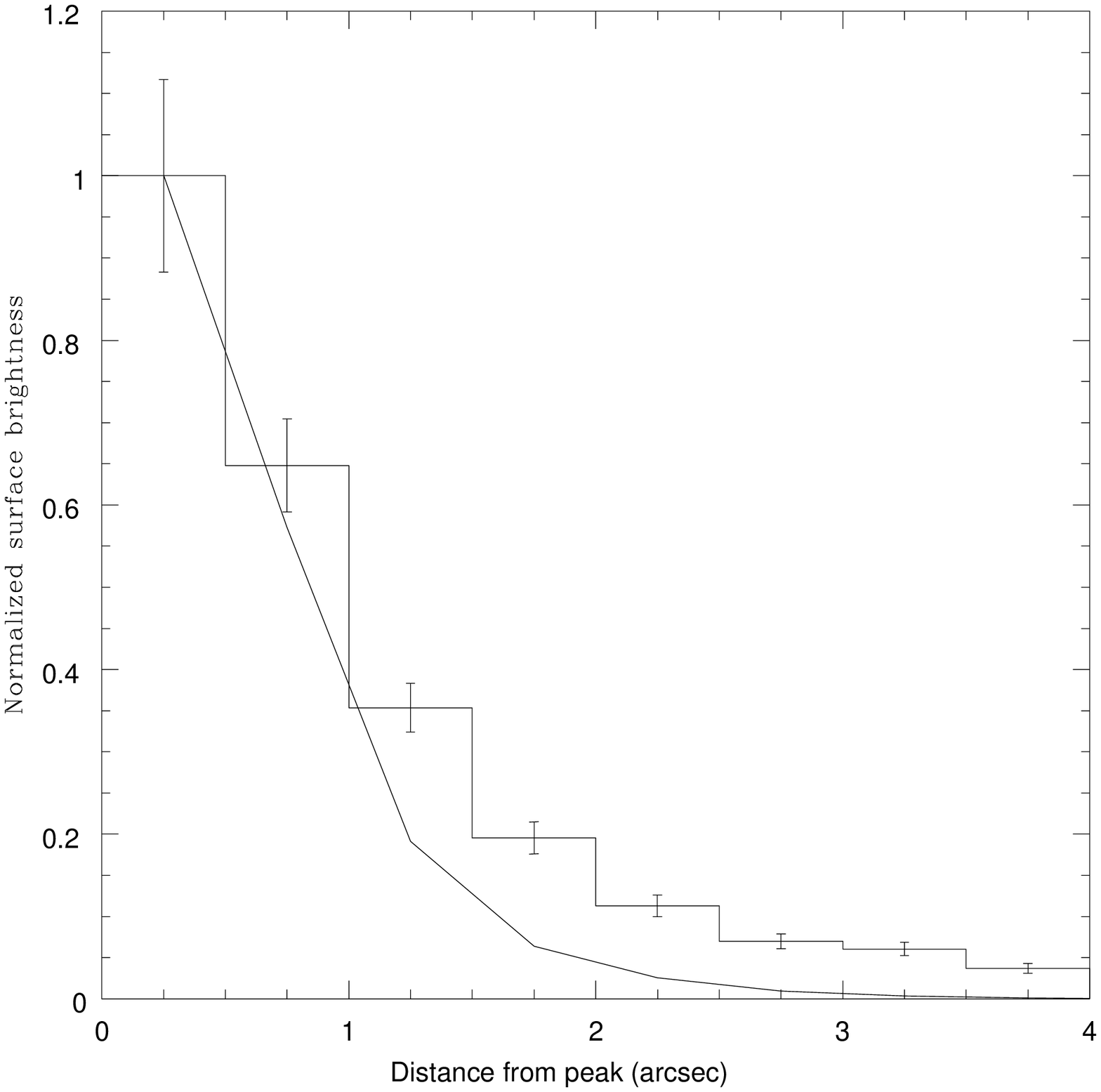}
\caption{Surface brightness (normalized cts arcsec$^{-2}$ s$^{-1}$)
distributions of X-ray knots AX1 (left) and BX2 (right) in $0.5''$ bins within
a $4''$ radius of the knot peak in the 0.4 to 2.5 keV band.  The surface brightness
distributions of the MARX simulated PSFs at the locations of the knots at 1.5 keV are shown
as the continuous curves.  The data have been normalized to have the unity surface
brightness in the first bin.  Errorbars are due to counting statistics only.}\label{sbp}
\end{figure}

To determine which of these knots are extended and which are
pointlike and may therefore be XRBs unrelated to the jet,
we have simulated point sources at five positions
separated by $1'$ along the jet starting at the nucleus using MARX.
We only used data from the first (OBSID 00316) observation as the
jet is closer to the best focus, and coadding the second
data set at a different roll angle would add considerable systematic uncertainty.
The ratio of the number of counts within
a $1''$ radius circle to the number of counts within a $1''$ to $2''$
annular ring was determined for each of these simulated point sources.
A third-order polynomial was fit to this ratio as a function of
distance from the nucleus in order to estimate this ratio
at an arbitrary position along the jet.
The measured ratio for each of the detected enhancements was then
compared to the simulated ratio.  If the measured ratio is
less than 3$\sigma$ below the simulated ratio, the enhancement
is considered a point source.  On the basis of this
test, we find that three of the enhancements
are consistent with point sources (see Table~\ref{knottab}).
The normalized surface brightness profiles of the two brightest knots,
AX1 and BX2, and the MARX simulations of the PSFs at the locations
of those knots is plotted in Figure~\ref{sbp}.
We have evaluated this procedure by testing other sources located
near the jet, but clearly outside the diffuse emission, and
find that four of the five sources are consistent with being point sources.  The fifth
source is obviously distorted and by visual inspection alone
it is clear that it should not be considered a point source.
The position, background-subtracted count rate, and luminosity of
each knot are given in Table~\ref{knottab}.
Based on the radial surface-density distribution of X-ray point sources in 
Cen A \citep{kra01}, we expect three of the knots or sources to be X-ray
binaries within Cen A, but unrelated to the jet.  
We conclude that the remaining knots are features of the
jet structure that are resolved by {\em Chandra}.

\begin{figure}
\plotone{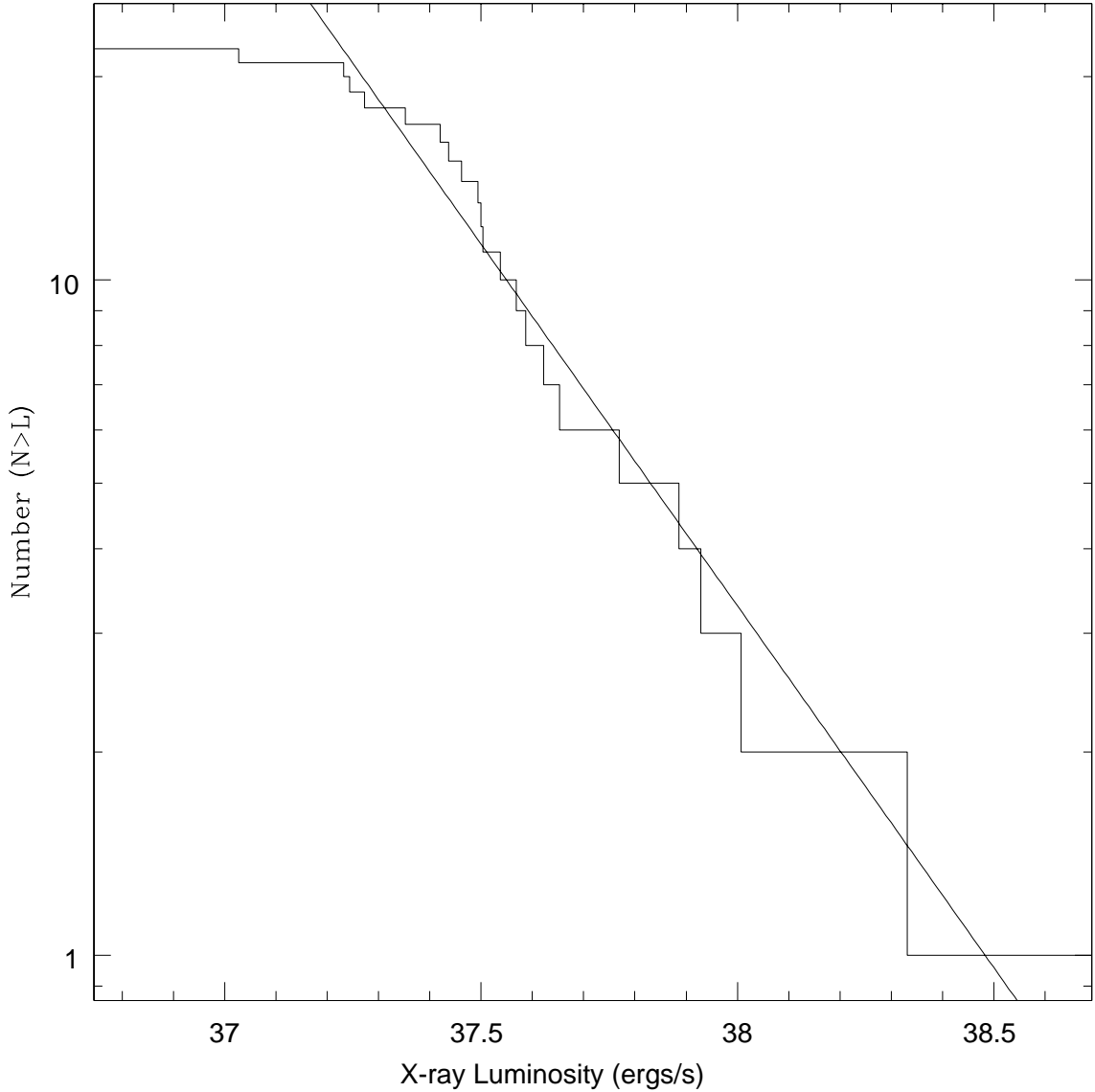}
\caption{Luminosity function of knots in X-ray jet, excluding the knots
of the inner jet and those identified as point sources.  The X-ray luminosity
is the unabsorbed luminosity in the 0.1-10 keV band assuming a power-law spectrum
with photon index 2.3 and $N_H$=1.7$\times$10$^{21}$ cm$^{-2}$.  The best-fit
power-law model is for $L_X>$3$\times$10$^{37}$ ergs s$^{-1}$ is also shown.}\label{klfunc}
\end{figure}

The X-ray morphology of the jet is clearly more complex when observed at
higher resolution.  It is therefore reasonable to expect that
there may be even more complex structure at smaller spatial scales
that can only be resolved at higher resolution.
To estimate what fraction of the remaining unresolved emission of the
jet could be due to a large number of unresolved, lower luminosity knots,
we created the luminosity function (LF) for the jet knots shown in Figure~\ref{klfunc}.
The luminosity of the knots is taken from Table~\ref{knottab}.  We assume
a power-law spectrum with photon index 2.3 and $N_H$=1.7$\times$10$^{21}$ cm${-2}$.
The knots of the inner jet (NX1 and AX1-6) were excluded
because the variable absorption makes determination of the
unabsorbed luminosity more uncertain.  The knots we
identified as point sources (CX1 and EX1) have also been excluded.
The limiting sensitivity for this analysis is somewhat uncertain because
of source confusion and depends
on the distribution of luminosity and spatial extent among the knots.
A detailed computation of the limiting sensitivity for detection of
knots in the jet would require a Monte Carlo simulation and is beyond
the scope of this paper.
We note that the limiting sensitivity away from the jet region
for which point source detection
is complete and unbiased is $\sim$3$\times$10$^{37}$ ergs s$^{-1}$ \citep{kra01}.
Source confusion and source extent will increase this value, but this increase is offset somewhat
because we coadded the two observations for this analysis.
For simplicity, we choose the limiting luminosity for knot detection, 
$L_{sens}$=3$\times$10$^{37}$ ergs s$^{-1}$, the same as for point source detection.
This is approximately the luminosity at which the LF begins to flatten
(see Figure~\ref{klfunc}).  The results below are not sensitive to this choice.

We fit a power-law to the knot LF for knots with $L_X>$3$\times$10$^{37}$ ergs s$^{-1}$ 
to quantify the distribution of knot luminosities.
We parameterize the LF as
$$ N(>L)=N_0\times (L/L_0)^{-\alpha},$$
where $N_0$ is the normalization at $L_0$=10$^{39}$ ergs s$^{-1}$.
The best-fit parameters are $\alpha$=1.1 and $N_0$=0.28.  If this
distribution is extended to some minimum luminosity, $L_{min}$, the integrated
X-ray flux from the jet (using the best-fit parameters) between $L_{min}$ and
$L_{max}$ (the luminosity of the brightest detected knot) is given by
$$L_T=3.09\times L_0(({{L_{min}}\over{L_0}})^{-0.1}-({{L_{max}}\over{L_0}})^{-0.1}).$$
The total luminosity of the jet in a region $195''$ long and $30''$ wide
beginning at a distance of $45''$ from the nucleus containing all the knots in the LF
is 3.59$\times$10$^{39}$ ergs s$^{-1}$.  The integrated luminosity of the
detected knots in this region is 1.23$\times$10$^{39}$ ergs s$^{-1}$.
The remainder of the emission of the jet can be entirely accounted for if 
$L_{min}$=3$\times$10$^{35}$ ergs s$^{-1}$, assuming there is no break
in the index of the LF below $L_{sens}$.  The total number of
knots at this minimum luminosity, $N(>L_{min})$, is $\sim$1700.
Based on this simple analysis, it is possible that the
unresolved diffuse emission of the jet actually consists of many lower
luminosity, and perhaps smaller scale, knots.
Each of the detected knots may, in fact, actually consist of several unresolved
lower luminosity knots.

\subsection{Comparison of X-ray and radio morphology}

\begin{figure}
\plotone{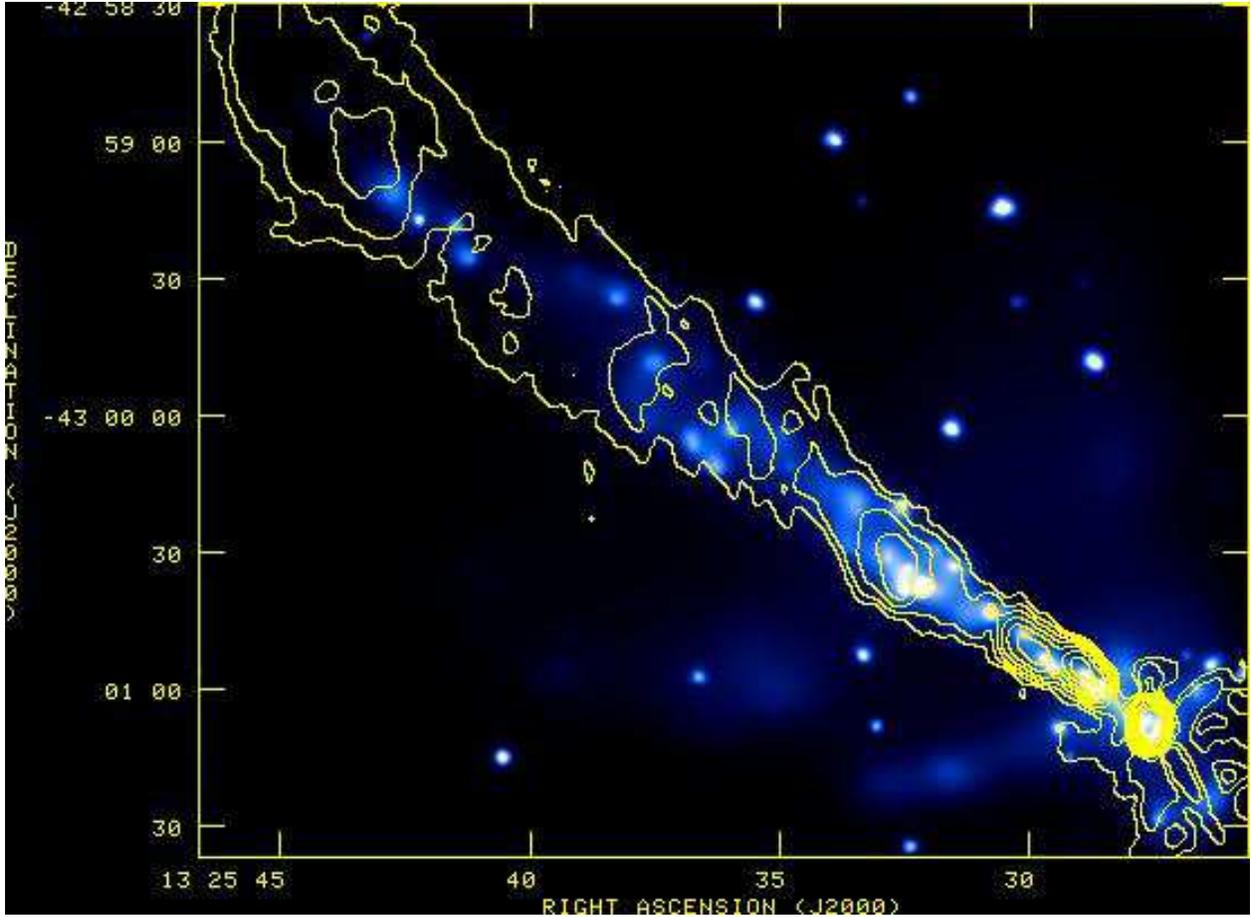}
\caption{Adaptively smoothed, coadded X-ray image in the 0.4-2.5 keV bandpass of the
jet in Centaurus A with 3.6 cm radio contours overlaid.  North is up and
east is to the left.  The radio beam is $3.39''$ (RA) $\times$ $4.70''$(DEC).}\label{jetrad1}
\end{figure}

\begin{figure}
\plotone{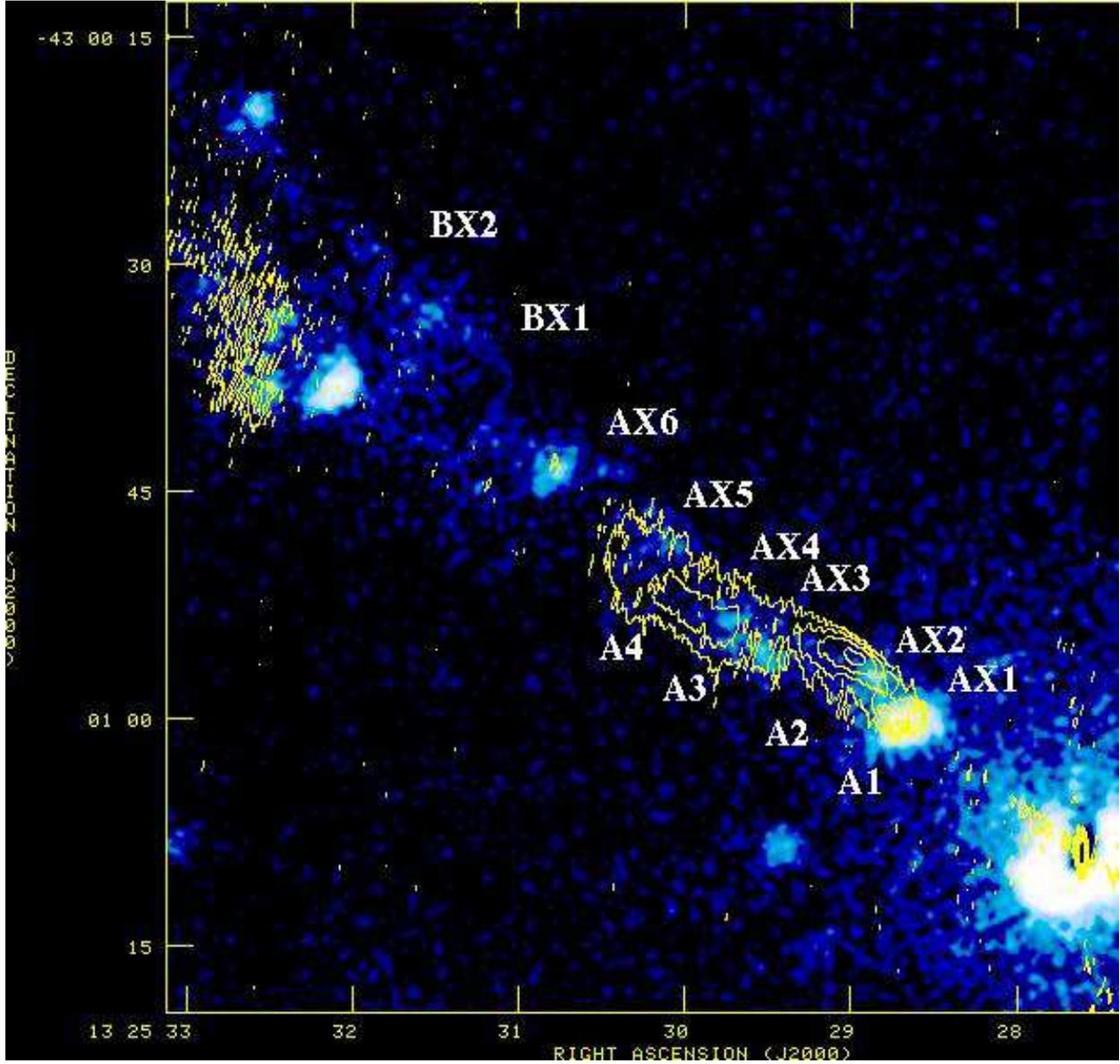}
\caption{Smoothed (Gaussian - FWHM=$0.5''$) X-ray image in the 0.4-5 keV
bandpass of the inner jet in Cen A with 3.6 cm radio contours overlaid.  North is up and
east is to the left.  The radio beam is $0.23''$ (RA) $\times$ $0.99''$(DEC).}\label{jetrad2}
\end{figure}

Figure~\ref{jetrad1} contains an adaptively-smoothed
X-ray image of the jet with 3.6 cm radio contours overlaid. 
The inner jet region, from the nucleus to just beyond
knot B, is shown in Figure~\ref{jetrad2} with the 3.6 cm radio
contours overlaid.  The resolution of radio data
is $3.39''$(RA)$\times 4.70''$(DEC) (FWHM) in Figure~\ref{jetrad1}
and $0.23''$(RA)$\times 0.99''$(DEC) (FWHM) in Figure~\ref{jetrad2}.
The approximate positions of the X-ray and radio knots are labeled 
in Figure~\ref{jetrad2} above
and below the jet, respectively.  
Within the $\sim 0.5''$
absolute positional resolution of the X-ray image, the position of radio knot A1
agrees with that of the first bright X-ray knot AX1.
We detect weak X-ray emission between the nuclear
jet (NX1) and this first bright X-ray knot
coincident with previously observed radio emission \citep{mar00}.
Further along the jet, however, there is a noticeable difference between
the peaks in the X-ray emission and the radio emission.
At knot B, the peak of the X-ray emission
is $5''$ ($\sim$80 pc) closer to the nucleus than
the peak of the radio emission.  The ratio of X-ray to radio flux 
for knots AX1/A1 and BX2/B is considerably higher than for the rest of the jet.  
Such an anomalous brightness at one or two knots
in the jets of several other FR I radio galaxies has
recently been reported \citep{hrd01}, and may be
a fundamental characteristic of such jets.

\begin{figure}
\plotone{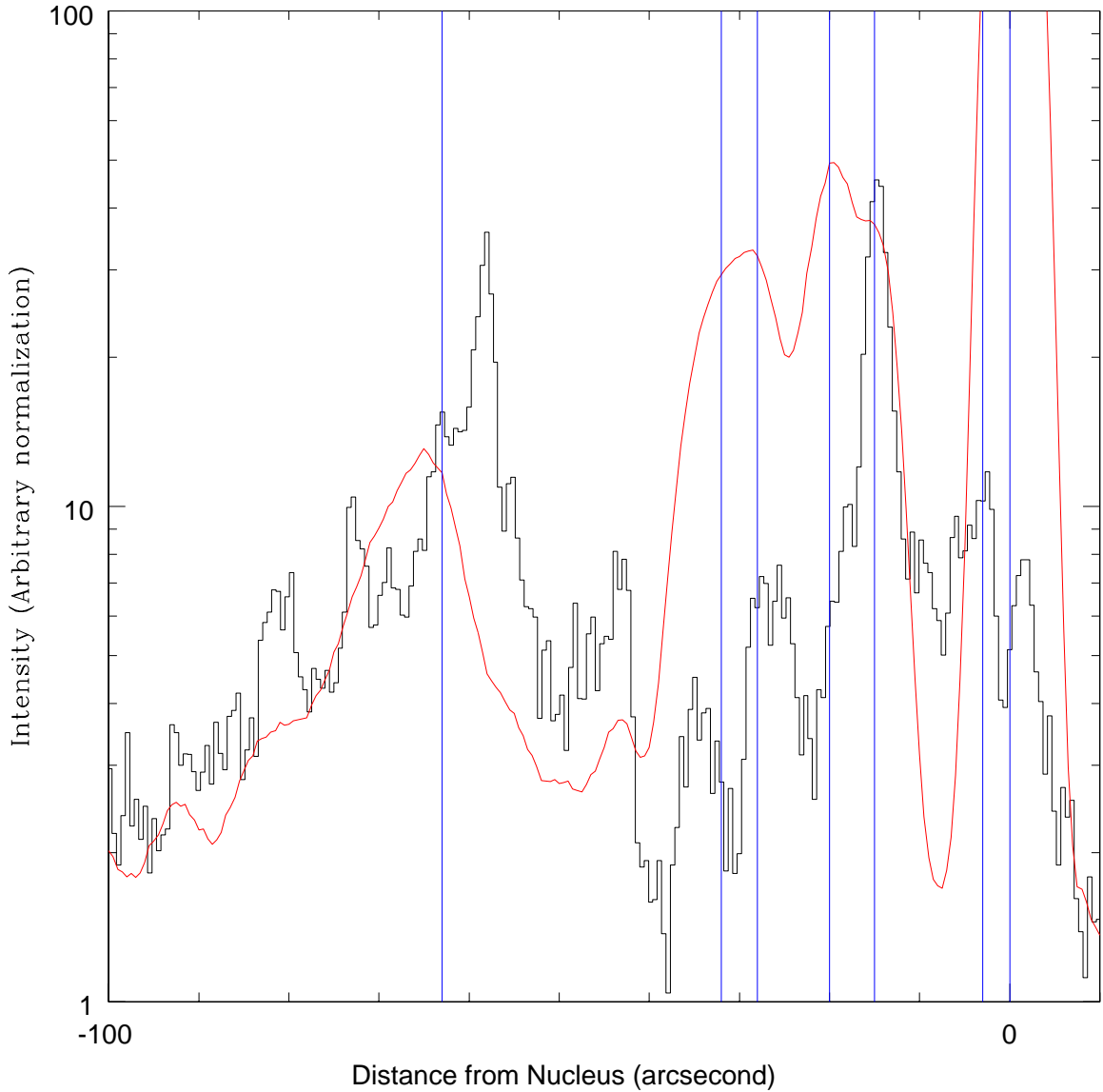}
\caption{Intensity profile (arbitrary normalization) of the X-ray
jet in the 0.4-1.8 keV bandpass along position angle 55${\mbox{$^\circ$}}$
in a region $24''$ wide (the black histogram).
The continuous curve (red) is the radio flux density at 3.6 cm (arbitrary
normalization) along the same projection.  The positions of the nucleus
and the radio knots N1, A1, A2, A3, A4, and B from previous radio
observations \citep{bur83} are shown as the blue vertical lines.
The slight difference in position of the peak of knot A1 is consistent with the
uncertainties in the absolute alignment.  The differences
in position between the peaks of X-ray and radio emission of the other
knots are clearly visible.}\label{profile}
\end{figure}

Although morphological differences between the X-ray
and radio emission are significant
both along the jet and perpendicular to it, these can be seen more clearly 
by viewing the jet in projection.  
Figure~\ref{profile} shows the intensity profile of the jet in a projection
$24''$ wide along position angle 55${\mbox{$^\circ$}}$
in both the X-ray (0.4-1.8 keV to suppress the mirror-scattered
nuclear contribution) and radio (3.6 cm).
This projection extends $100''$ from the nucleus along the
jet to a region just beyond knot B.
No background has been subtracted from the X-ray data but, as
can be seen from the gaps between the knots, the background
is small in this projection.
The radio positions of the knots (A1-4 and B) \citep{bur83} of the inner jet have
been labeled.  Differences and offsets between the X-ray and radio emission of
the jet are apparent.  There is a small difference between the
radio and X-ray peaks at knot A1/AX1, but this difference is
consistent with our $\sim 0.5''$ uncertainty on the absolute alignment of
the X-ray data and centroiding the position of the X-ray nucleus
because of the pile-up.
Beyond knot A1, however, the morphological differences are
significant.  The second (AX2) and third (AX3) X-ray knots of the inner jet
lie approximately $2.5''$ ($\sim$ 40 pc) closer to the nucleus than the second (A2)
and third (A3) radio knots, whereas the fourth knot
(AX4) lies approximately $5''$ farther from the nucleus than
the fourth (A4) radio knot.  Knot AX6 (knot AX in \citet{kra00})
coincides with a small peak in the radio emission.
The region between the fourth radio knot (A4) of the inner
jet and radio knot B is crossed by several
dark optical bands (see Figure~\ref{optovl}).
The radio position of the nuclear jet is several arcseconds
from the nucleus \citep{bur83} and is labeled (N1).
We detect a significant X-ray enhancement at the position of
this nuclear jet (labeled NX1), and continuous emission from this knot to
X-ray/radio knot A1.  The X-ray emission from knot B is composed of emission
from one bright knot, which lies $\sim4''$ closer to the nucleus
than the radio peak, and five fainter knots surrounded
by diffuse emission.  

These high resolution {\em Chandra} observations permit us
to measure the transverse width of the X-ray
jet as a function of distance from the nucleus.
As can be seen from Figure~\ref{jetbw},
the jet progressively widens along its length, and several bends or turns
are visible in the inner jet region between the nucleus and knot B.  In the inner jet
region, the width of the jet is only marginally resolved in our
data.  As described above, we used MARX simulations
\citep{wis97} to model the PSF of the observation at 1.5 keV for the location of
the inner jet in the first observation (OBSID 00316).  
We used only one observation for this analysis
to avoid uncertainties that would be introduced
by combining the PSFs of two observations at different roll
angles and off axis positions.  
In the inner jet region, we estimated the width of the jet
perpendicular to the direction of propagation according to
\begin{equation}
\sigma _{knot}=\sqrt{\sigma _{obs}^2-\sigma _{psf}^2},
\end{equation}
where $\sigma _{knot}$ is the width of the knot (FWHM), $\sigma _{obs}$ is
the observed width (FWHM) of the knot, and $\sigma _{psf}$ is the width of
the PSF (FWHM).  Beyond the inner jet, the transverse width of the
jet is clearly resolved and can be measured directly.

\begin{figure}
\plotone{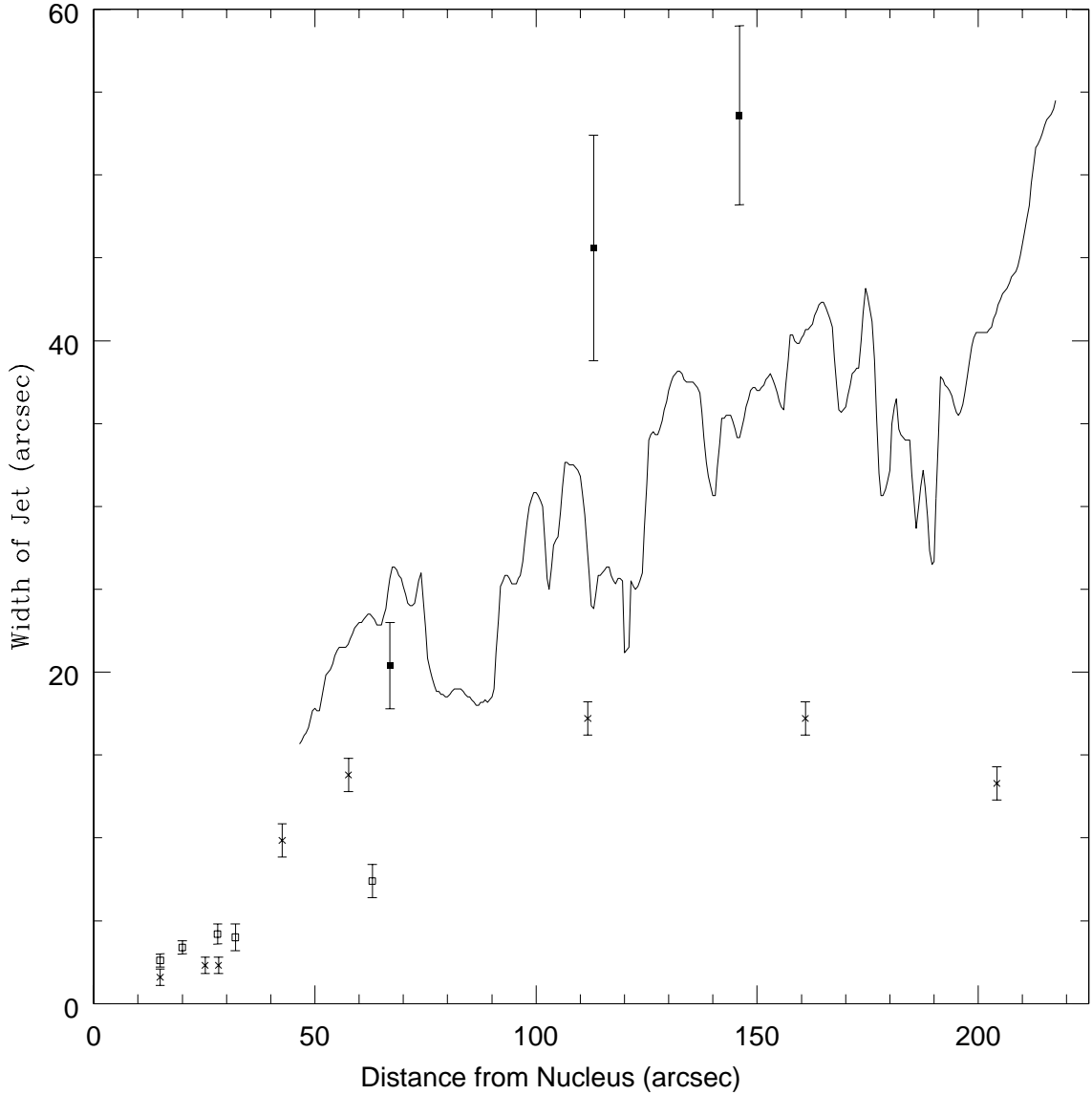}
\caption{Transverse width of jet as a function of distance from
the nucleus.  The X's are the width of the X-ray jet (see equation~1) measured
from our {\em Chandra} observations,
and the open and solid squares are the widths of the jet at 20 cm at
high and low resolutions, respectively, taken from Table~2 of \citet{bur83}.  The
continuous curve is the width of the jet at 3.6 cm measured from the
data shown in Figure~\ref{jetrad1}.}\label{width}
\end{figure}

The width of the X-ray jet as a function of distance from the nucleus is
plotted in Figure~\ref{width}.  For comparison,
the FWHM of the individual features in the jet
taken from earlier radio measurements (taken from Figure 4 and Table 2 of
\citet{bur83}) at two angular resolutions
are also plotted.  The width measured from our more sensitive
3.6 cm radio data shown in Figure~\ref{jetrad1} is plotted as the continuous curve.
We note that these are two somewhat different definitions of width.
\citet{bur83} define the radio width of the jet as the second moment
of the brightness distribution.  We have chosen the surface-brightness boundary
of the jet as the definition of width for the 3.6 cm data shown as the continuous
curve in Figure~\ref{jetrad1}.  This difference permits a more straightforward comparison
with our X-ray data as we have chosen two similar definitions
of the width for our X-ray analysis.  It is appropriate to compare
the radio widths of \citet{bur83} with our measurements of the X-ray
widths in the inner jet region, and the 3.6 cm radio widths with the
X-ray widths beyond the inner jet region.
As can be seen from Figure~\ref{jetrad1}, the X-ray width of the
jet at knot AX1 is marginally smaller than the radio width.
Beyond knot AX1, however, the X-ray jet is definitively narrower than the radio jet.
It is not clear whether this is a real feature or an
instrumental artifact related to the low X-ray surface brightness at
the edge of the jet as it becomes undetectable above the emission from the hot ISM.
If this difference in widht is a real feature of the jet, it implies that the higher energy,
X-ray emitting plasma is confined in a narrower channel than the radio emitting plasma.

\begin{figure}
\plottwo{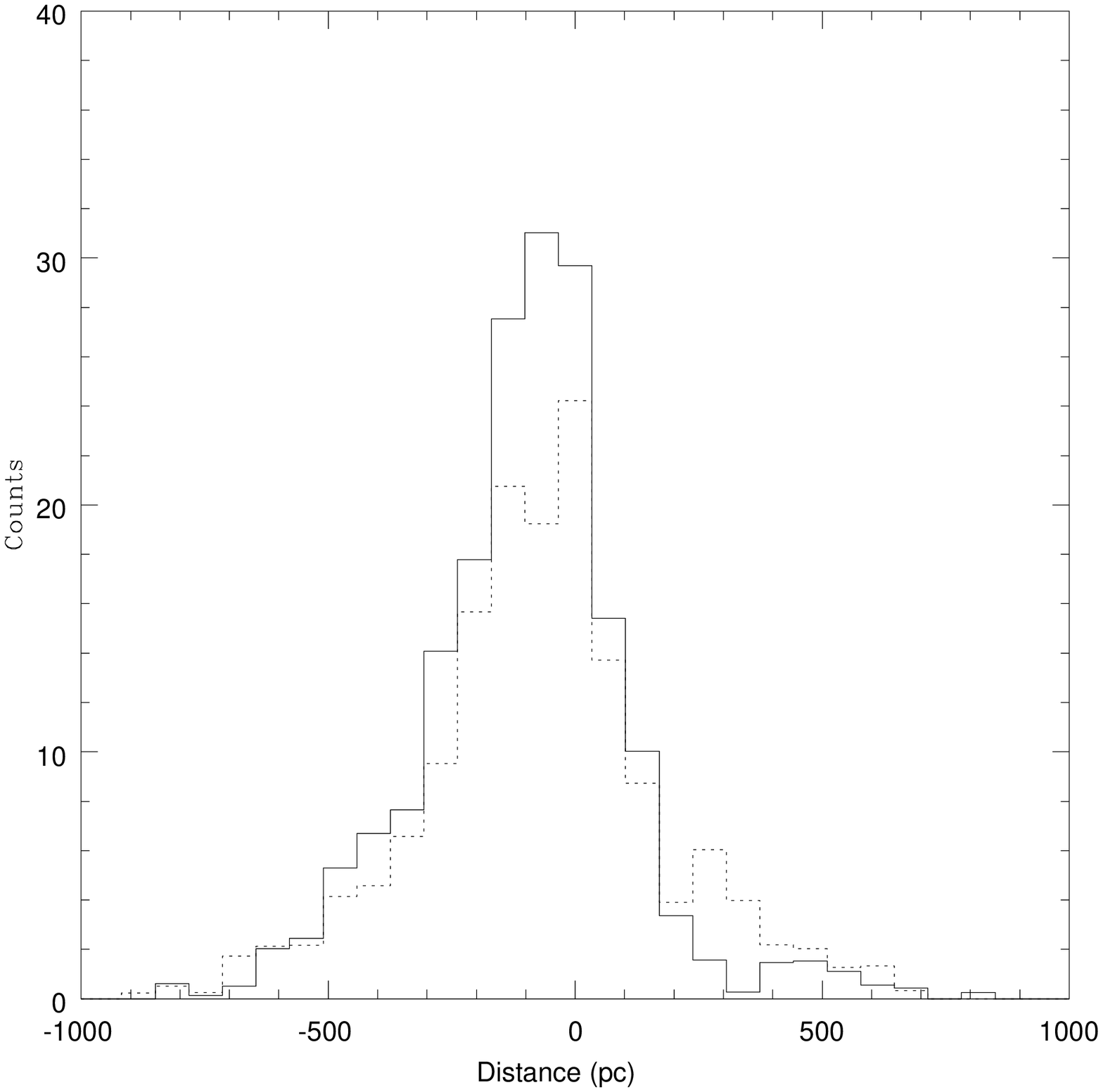}{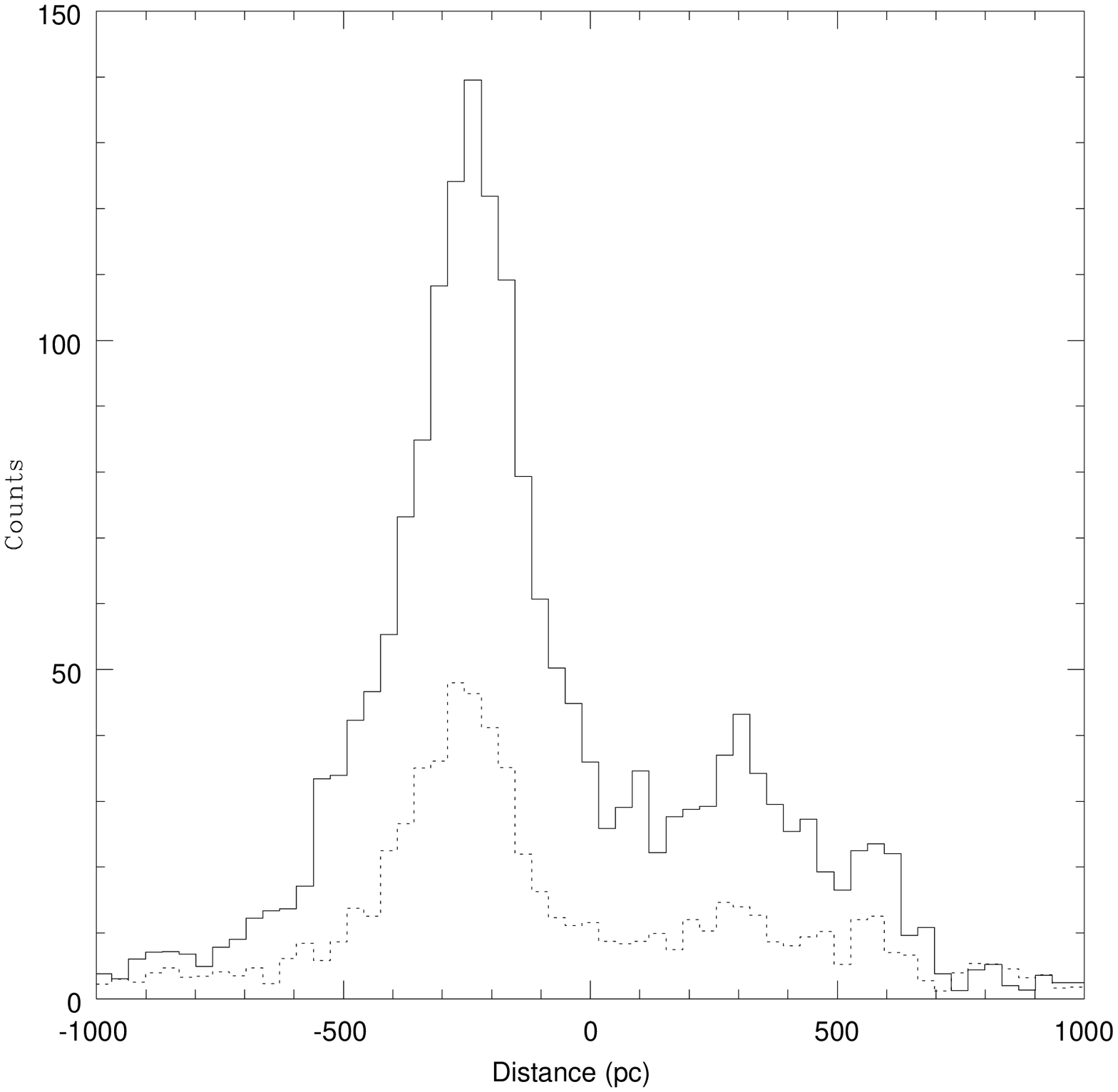}
\caption{Projection of the jet along position angle 55$^\circ$ showing the
transverse width in two energy bands: 0.4-1.5 keV (the solid line)
and 1.5-5.0 keV (the dashed line).  The data from the two
observations have been co-added.  The left hand figure is
the projection of the region between knots AX1 and AX4, inclusive,
and the right hand figure is the knot B
region.  See Table~\ref{projregs} for definition of the
regions. For a given region, the width of the jet in the
two energy bands is the same.  The difference in relative normalization in the two energy
bands between the two regions shown here is due to the larger $N_H$ of the
first region.}\label{proj1}
\end{figure}

We also made projections of the data along a position angle
of 55 degrees in two different energy bands, 0.4 - 1.5 keV
and 1.5 - 5.0 keV to investigate whether there are differences
in the width of the jet in the two X-ray bands which would
support the idea that the most energetic particles are
constrained within a smaller channel of the jet.  Four regions
of the jet were projected; the inner jet between, but not
including, knot AX1 and knot B, knot B, and two regions beyond
knot B.  The regions are summarized in Table~\ref{projregs}.
Figure~\ref{proj1} contains the first two of these projections;
the other two are similar.  The data from the two observations have
been co-added but not exposure corrected.  We find no statistically significant
difference in the width of the jet in the two bands
at any position along the jet implying that the particles
responsible for the emission in the two bands are
homogeneously mixed.  This result also suggests that variable
absorption is not significant in modifying the transverse
morphology of the jet.

Two knots of radio emission in the SW radio lobe
\citep{cla92} could be related to a counterjet.
Similarly, several pointlike sources of X-ray emission have been
detected in both the HRC image \citep{kra00} and the ACIS observations presented
here.  There are five bright X-ray points
that are approximately colinear with the nucleus and the
forward jet, suggestive of a counterjet.
Several of these points are near ($\sim5''$), but not
coincident with the radio knots.
The two radio knots are not colinear with the nucleus and forward jet.
X-ray spectra of these unresolved sources are not significantly different
from that of the population of X-ray binaries \citep{kra01}.
It is possible that the emission from
these unresolved X-ray sources may be related to a non-thermal process
originating in a collimated outflow from the nucleus,
but without additional evidence, we conclude that these
sources are XRBs within Cen A unrelated to a counterjet.

\subsection{Spectral Analysis}

We derived spectral parameters for four regions
of the jet.  These regions are knot A1, knot B, the knots between A1 and B, and all of
the emission beyond knot B to the edge of the X-ray jet.
It was not possible simply to combine the data from the two observations for
spectral analysis for two reasons.  First, the focal plane
temperature at the time of the first observation was -110$^\circ$C,
but it had been changed to -120$^\circ$C by the time of the
second.  Second, even though the entire jet was placed on one
chip for both observations, the jet spans different parts of
the chips for each.  Given the variation in spectral response
across output nodes and across a chip due to radiation damage, it was
necessary to generate response matrices for each observation and
fit data from both observations simultaneously.  For the last two regions,
a response matrix appropriate for the center of
the given region was used.  None of the regions
crosses output node boundaries.  

We have fit absorbed power-law, Raymond-Smith with variable abundance,
and thermal-bremsstrahlung models to the spectra in the 0.4 to 4.5 keV band.
All three models produce acceptable fits, and
there is no statistical reason to choose one over the other.
The best-fit temperatures of the Raymond-Smith and thermal-bremsstrahlung
models range between 3 and 5 keV with large uncertainty,  The abundance in the
Raymond-Smith fits is low ($\sim$0.1 solar but statistically
consistent with zero) and poorly constrained.
Such a high-temperature, low-abundance plasma has no spectral
features or emission lines than can be resolved at the resolution
of the ACIS-I instrument and is indistinguishable from the
absorbed power-law model when fit over a restricted energy band.
Although our data do not rule out a thermal model,
we reject them on physical grounds as described below and consider only
the absorbed power-law model further.

\begin{figure}
\plottwo{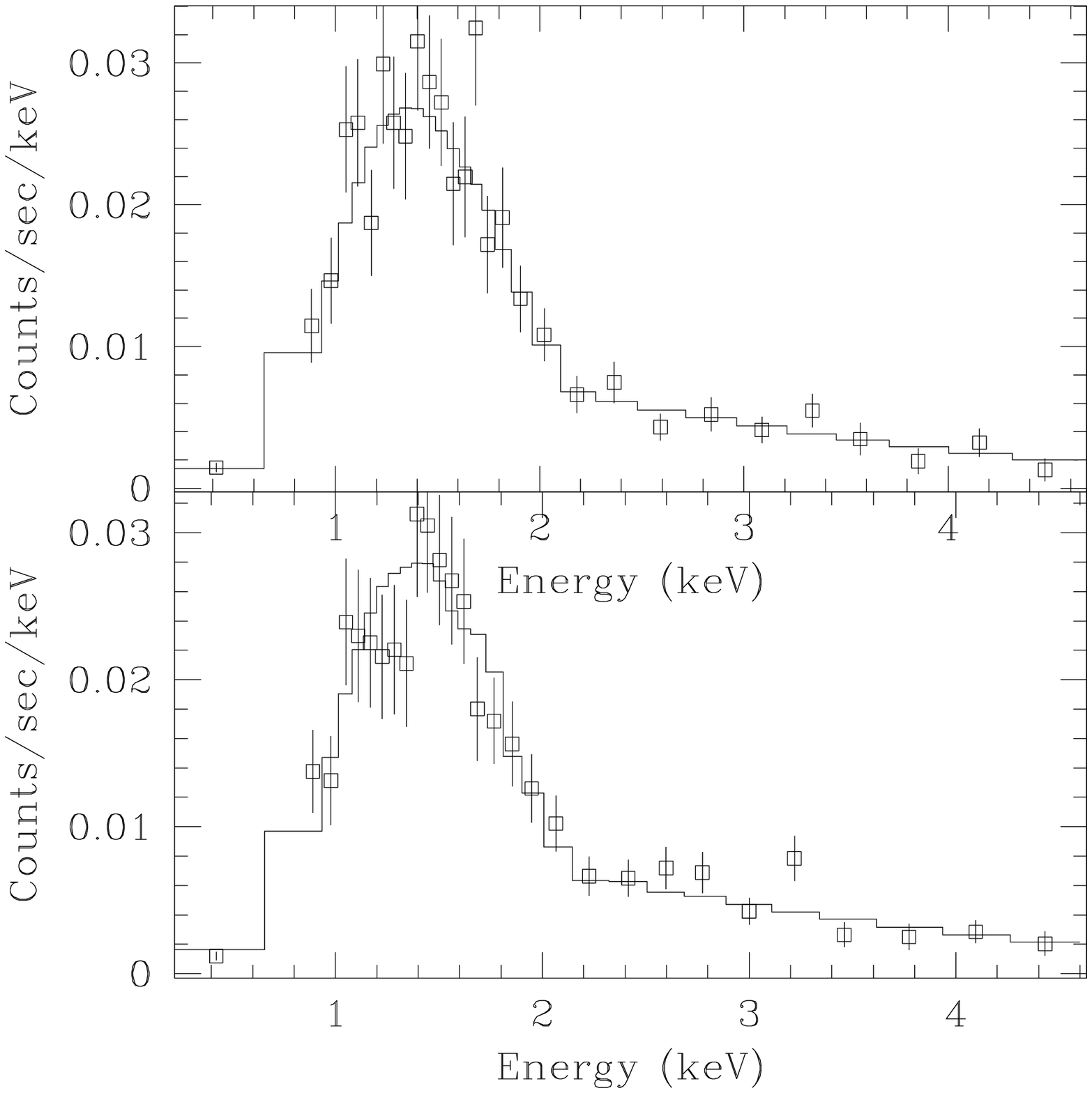}{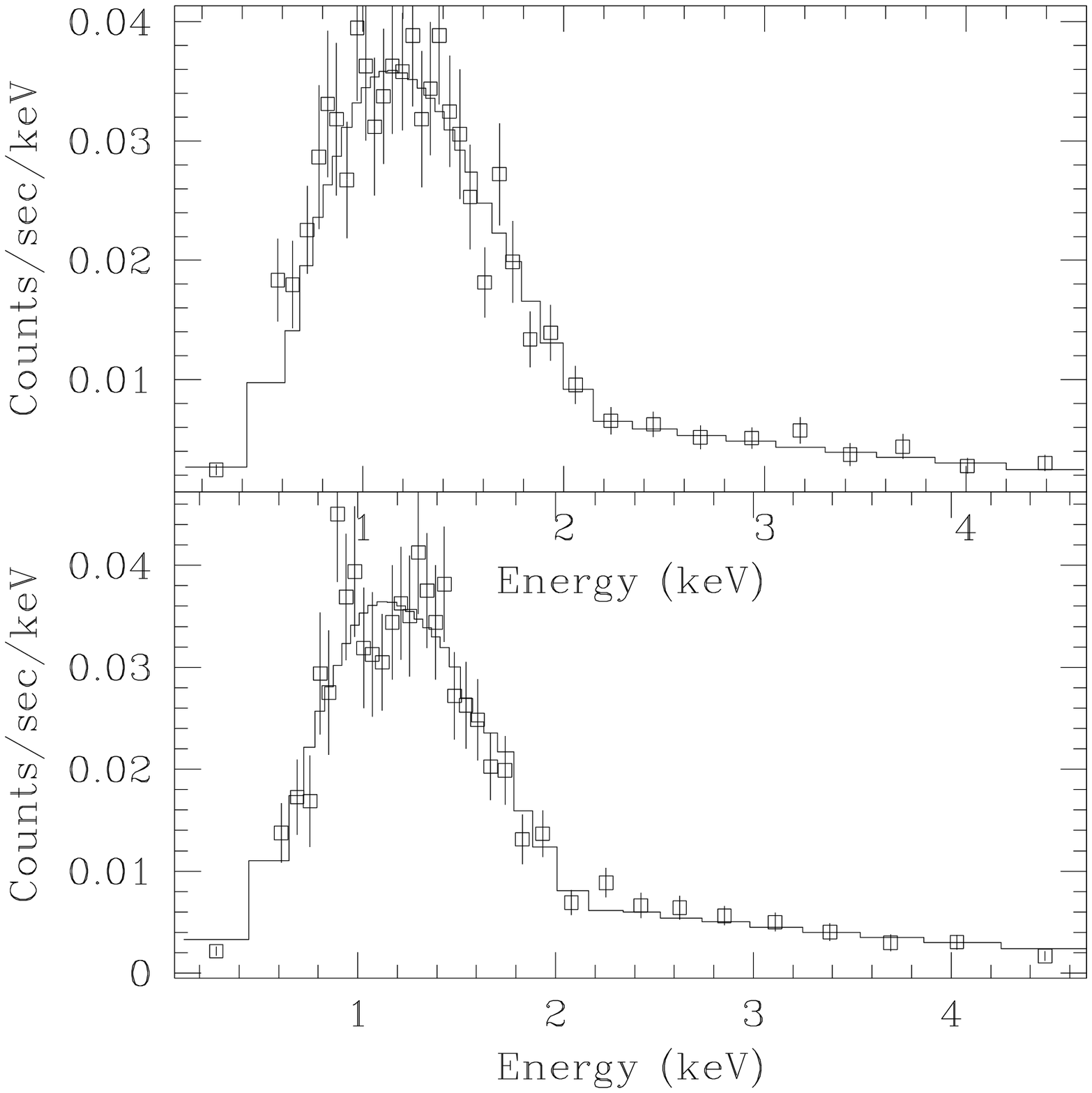}
\caption{Spectra and best-fit absorbed power-law model of knot AX1/2 (left)
and knot B (right).  The spectra on the top are the data from the
first observation (OBSID 00316), and those on the bottom from the second (OBSID 00962).  The best
fit parameters are summarized in Table~\ref{sfit}.}\label{knotspect}
\end{figure}

The spectra and best-fit power-law for knots A1 and B are shown in Figure~\ref{knotspect},
and the parameters of all of the fits of the absorbed power-law model 
are summarized in Table~\ref{sfit}.
Note that in all cases the best-fit absorbing column is larger
than the Galactic value of 7$\times$10$^{20}$ cm$^{-2}$ \citep{sta92}, implying
considerable absorption within Cen A.  This is not surprising
for the inner jet region where the jet crosses the dust lane.
As described above, the best-fit column density for the two inner jet regions
is at the high end of the range suggested by $A_V$.  Significant amounts of neutral hydrogen
($N_H\sim$ few$\times$ 10$^{20}$ cm$^{-2}$) have been detected in the vicinity of
the jet beyond knot B out to knot F \citep{sch94}, implying that perhaps the jet
is behind much of this absorbing material.
The best-fit index for the emission beyond knot B is similar to that
of the brighter knots of the inner jet.  The flux from this region of the
jet is dominated ($\sim$75\%) by the diffuse, unresolved component.  This
spectral similarity between the knots of the inner jet and the diffuse
component further from the nucleus implies that the emission mechanisms
are similar, and supports the hypothesis, suggested in the LF analysis
above, that the the unresolved, diffuse emission is composed of many
lower-luminosity, spatially unresolved knots.

\subsection{Radio to X-ray Spectral Index}

We have determined the radio to X-ray spectral index for each of the
bright radio and X-ray knots in the inner jet.  Even though the
X-ray and radio peaks are misaligned, there is detectable X-ray emission between
the radio knots and vice versa.
We measured the X-ray count rate and radio flux
density at 8.4 GHz in $4''$ wide boxes centered
on each of seven X-ray and three radio peaks of the inner jet and in the vicinity
of radio knot B.  The X-ray count rates were converted to flux densities at 1 keV
assuming a photon index of 2.2.
For the X-ray knots (excluding knot AX1),
we find that the spectral indices between the X-ray and radio bands
vary between 0.80 and 0.97 (average 0.90), and for the radio knots (again
excluding AX1) between 0.99 and 1.01.
The flux densities and spectral indices 
are summarized in Table~\ref{fdens}.
Even though there are significant (factor of three)
variations in the ratio of X-ray to radio
flux along the jet (Figure~\ref{profile}), the differences
in the spectral index between the X-ray peaks and
radio peaks are only $\sim$0.1-0.15.
All of these indices are considerably steeper than the radio
spectral indices (20cm to 6cm) previously reported \citep{fei81},
implying spectral steepening between the radio and the X-ray.
We note that these radio to X-ray indices are remarkably
similar to those measured for the jet in the FR I galaxy 3C 66B \citep{hrd01},
but considerably steeper than that of 3C 273 \citep{mar01a,sam01}.

\section{Discussion}

To summarize, we find that
the X-ray morphology of the jet consists of an unresolved, "diffuse"
component that extends from within at least $\sim$50 pc ($\sim 3''$)
of the nucleus to $\sim$4 kpc, and contains at least
31 distinct, spatially-resolved X-ray enhancements embedded within
diffuse emission.  The width of the diffuse component of the inner jet is roughly
constant to knot AX6, where it widens, although
there appears to be a bend or twist in the
jet between knots AX2 and AX3 in this inner jet region.  In the
vicinity of knot B, the jet widens again.  Beyond knot F the jet appears to
narrow, but this may be due to a decrease in the surface
brightness of the diffuse component.
There are significant differences between the X-ray and radio
morphologies in the inner jet region.
The spectra of the knots and diffuse emission of the jet
are well fit by both absorbed power-law and absorbed thermal
models, but there is no direct evidence for any emission lines
in the spectra.  What do these results tell us about the nature of the X-ray
emission?

Generally, four models are typically invoked to explain
X-ray emission from extragalactic jets and the hotspots in
radio lobes: thermal emission, synchrotron
self-Compton (SSC) emission, inverse-Compton scattering
from an external source of seed photons, and
synchrotron emission from a population of ultra-relativistic electrons \citep{har01}.
The first three have been rejected by previous authors
\citep{fei81,bur83} for the Cen A jet, and this work does not alter
those conclusions.  The thermal model was ruled out because of the relatively
high density of thermal plasma required to explain the
emission.  This plasma would be overpressured relative to the
surrounding medium, and it would therefore be difficult to confine this plasma
within a small volume.  The knots would rapidly dissipate.
In addition, Faraday rotation and depolarization measurements
put upper limits on the density of material in such
thermal knots that are well below that required to account
for the observed emission.
The absence of any significant spectral features observed in our data
strengthens this conclusion.  We note that recent {\em Chandra}/HETG observations of
the jet of M87 also failed to detect line emission \citep{mil01}.
Inverse-Compton scattering of the Cosmic Microwave Background
has been successful in explaining the X-ray emission from more
powerful jets such as 3C 273 \citep{sam01} and PKS 0637-752 \citep{cel01},
but only by invoking a large bulk Lorentz factor for a jet placed
at small angle to the line of sight. It is believed, however, that the Cen A jet lies
at a large angle (50$^\circ$-80$^\circ$) to our line of sight \citep{jon96,isr98,tin98}.
The SSC mechanism can be rejected as the density of
radio photons is even lower than CMB photons, making any such model even more
unlikely.  Another possible source of seed photons is beamed optical emission 
from the active nucleus \citep{per85}.  In the
unified AGN model, Cen A is considered to be a misdirected BL Lac \citep{urr95},
so the beamed emission from the nucleus, unseen by us, could be quite
large \citep{fos98}.  Electron energies of $\gamma\sim 25$ are
required to boost optical photons into the X-ray regime.  At knot
A1, the synchrotron frequency of these low energy electrons is below
1 MHz (assuming the equipartition magnetic field of 61 $\mu$G), which is unobservable.
The observed X-ray spectral index (energy index 0.9-1.4) of the inner jet, however,
is much steeper than the radio spectral index (0.4-0.6) between 1.4 and 4.87 GHz.  Unless the
spectral index of the unseen electrons is dramatically steeper than
that of the electrons producing the observed radio emission,
it is not possible for a significant fraction of the
X-ray emission to be due to this mechanism because of these spectral differences.

On the basis of the morphological similarities between the radio and X-ray emission
and the limitations of competing models,
it has previously been concluded that both the X-ray and radio flux from the
jet originate in synchrotron emission from a population of ultrarelativistic electrons.  
The knots are presumably sites of shocks where particles
are accelerated to $\gamma\sim$ 10$^{7-8}$ to produce
the observed X-ray emission, assuming an equipartition
magnetic field \citep{fei81,bur83}.
The main argument against a synchrotron
hypothesis has been that the lifetime due to synchrotron emission of the
ultra-relativistic particles is considerably less than
the travel time down the jet, thus requiring that the
particles be reaccelerated \citep{fei81}, perhaps
multiple times, as they travel along the jet to the radio lobe.
In fact, in the inner jet region, the particles must
undergo local accelerations multiple times within a knot.
The lifetime of the X-ray emitting electrons in the inner jet
is $\sim$20 years, an order of magnitude less than the
travel time from the nucleus to knot B.   The lifetime
of the particles beyond knot B is longer ($\sim$60 years)
due to the lower magnetic field, but still considerably
less than the travel time into the radio lobe.
Based on estimates of the knot sizes from {\it Einstein}
observations (hundreds of parsecs), it was argued that the large volumes of the
knots pushed acceleration models to their limits \citep{bur83}.
This criticism is considerably weakened based on our {\em Chandra}
observations because each previously observed X-ray knot is
actually composed of several smaller knots. 
Thus the knot volume and the shock sizes are considerably smaller (tens
of pc rather than hundreds of pc) than previously supposed.
Local acceleration of the electrons within the knots is
still a necessary component of this model, however.

The morphological differences between the
X-ray and radio emission complicate this simple model for the emission .
These differences argue strongly against
the simple hypothesis that a single, spatially coincident, population of relativistic electrons
is responsible for both the X-ray and radio emission \citep{fei81,bur83}
in the nuclear and inner jet regions, except perhaps at knots N1/NX1 and A1/AX1.
Our data support a model in which the X-ray emission is
indeed synchrotron radiation from relativistic particles, but the
radio emission originates, at least in part,
from a spatially distinct, less energetic population of electrons.

The only other extragalactic jet that has been as well studied
as the Cen A jet is that in M87, and there are some
remarkable morphological similarities between the two jets.
Small differences between the X-ray, optical, and radio morphology
of the jet in M87 on roughly the same spatial scale as those described here
have been reported \citep{neu97,spa96,per98}.
In particular, recent {\em Chandra} observations of the M87 jet
have detected morphological differences remarkably similar to
those seen here \citep{per01,mar01b}.
HST imaging and polarimetric observations of M87 have provided evidence
that the more energetic particles responsible for
the optical and X-ray emission from the M87 jet are located closer
to the axis of the jet which is surrounded by a sheath or
cocoon of lower-energy particles responsible for the radio
emission.  The knots in the M87 jet are identified as the sites
of shocks in the flow where the magnetic fields are
compressed and the particles accelerated \citep{spa96,per98}.
The optical and X-ray emission is then due to
synchrotron radiation at the sites of the shocks.
Differences between the optical and radio emission are
caused by particle diffusion and aging.
A similar model can be used to qualitatively describe all the
features of the bright X-ray/radio knots of the Cen A inner jet.
The differences in the positions of the knots
in the X-ray and radio emission are naturally explained by particle aging.  
This hypothesis could be greatly strengthened if optical
emission could be detected between the various radio and
X-ray knots, but the dark dust lane makes such a detection
unlikely \citep{mar00}.
In fact, it is now becoming clear that such
morphological differences are a common feature of X-ray and
radio emission from FR I galaxies \citep{hrd01}.
The existence of shock sites and the generation of X-ray emitting plasma 
may be a fundamental feature of jets in FR I galaxies.

\section{Summary and Conclusions}

We have presented high-resolution {\em Chandra}/ACIS-I
X-ray images and spectra of the X-ray jet in Centaurus A and have found the
following:
\begin{enumerate}
  
  \item In the inner arcminute of the jet,
there are significant morphological differences between
the radio and X-ray emission.
The X-ray and radio positions of the two knots closest
to the nucleus (NX1 and AX1) agree to within the
precision of our absolute astrometry ($0.5''$), but beyond
knot AX1 ($\sim$250 pc from the nucleus), there are
significant differences up to $\sim$90 pc between the positions of
the X-ray and radio peaks.

  \item Given the offsets of the radio and X-ray knots,
there are large (factor of $\sim$3) variations in the ratio of
X-ray to radio flux in the inner jet region.
In the inner jet and the vicinity of radio knot B,
the radio to X-ray spectral indices of the X-ray
bright knots and radio bright knots
are about 0.9 and 1.0, respectively.

  \item At a linear resolution of $\sim$30 pc ($1.5''$) the
jet contains at least 31 knots or enhancements in X-ray emission
embedded within continuous diffuse emission extending
about 4 kpc from the nucleus.  Each previously
observed X-ray knot is composed of several
knots embedded within diffuse emission.

  \item The spectra of several regions of the jet are well fit
by an absorbed power-law model, and although we cannot
exclude thermal models, there is no statistically
significant evidence for any emission lines.
The spectrum of the jet beyond knot B, which is dominated
by the diffuse, unresolved component, is similar to that of
the bright knots of the inner jet.

  \item At the bright X-ray knot AX1, the width of the jet in the X-ray and radio
bandpasses is similar.  Beyond this knot into the NE radio lobe,
however, the X-ray jet appears to be narrower than the radio jet.
The X-ray jet appears to be well collimated from the nucleus
into the NE radio lobe.
\end{enumerate}

These results strengthen the case that the X-ray emission from
the Cen A jet is due to synchrotron radiation from ultra-relativistic
particles, and that the knots are shock sites where
the particles are re-energized as they travel along the jet.
There are strong arguments against thermal and inverse-Compton models for
the X-ray emission, none of which has been negated by these observations.
The strongest arguments against the synchrotron hypothesis were
the large shock volume, the difficulty of particle re-energization, and
the necessity of multiple re-acceleration within a knot.  These
have been significantly weakened by this observation as
the sizes of the knots are considerably smaller (tens of
pc rather than hundreds) than previously supposed.  
In addition, both the spectra and the radio/X-ray morphological differences we observe
are consistent with the synchrotron hypothesis.

{\em Chandra} has now detected X-ray emission from the jets
of a significant number of FR I radio galaxies \citep{wor01,har01}.
The radio/X-ray morphologies, the spectral indices and the
X-ray spectra of these objects are remarkably similar to those of Cen A.
We suggest that X-ray emission may, in fact, be a common aspect of
the jets in these galaxies, and that the shock
acceleration of particles to $\gamma\sim$10$^{7-8}$ may
be an integral feature of the hydrodynamics of such jets.
{\em Chandra} will allow us to make a systematic study of the X-ray emission
from extragalactic jets to better assess the relationship between bulk flows and
particle acceleration.

\acknowledgements

This work was supported by NASA contracts NAS8-38248, NAS8-39073, the
Chandra Science Center, and the Smithsonian Institution.
We are grateful to Jack Burns for allowing us to use his archival VLA
observations of Cen A.  We would also like to thank Herman Marshall
for discussing the results of the {\em Chandra} observations of the
M87 jet before publication.

\clearpage

\clearpage

\clearpage

\clearpage

\clearpage

\clearpage

\clearpage

\clearpage

\clearpage

\clearpage

\clearpage

\clearpage

\clearpage

\clearpage

\clearpage

\clearpage

\begin{table}
\begin{center}
\begin{tabular}{|l|c|}\tableline
Radio Knot & Distance from Nucleus (arcmin) \\ \tableline
N1 & 0.05 \\ \tableline
A1 & 0.25 \\ \tableline
A2 & 0.33 \\ \tableline
A3 & 0.47 \\ \tableline
A4 & 0.53 \\ \tableline
B  & 1.12 \\ \tableline
E  & 1.88 \\ \tableline
F  & 2.43 \\ \tableline
X-ray Knot & Distance from Nucleus (arcmin) \\ \tableline
A & 0.23 \\ \tableline
B & 1.00 \\ \tableline
C & 1.25 \\ \tableline
E & 1.85 \\ \tableline
F & 2.27 \\ \tableline
G & 3.40 \\ \tableline
\end{tabular}
\caption{Position of radio and X-ray knots of Centaurus A jet taken 
from \citet{bur83} and \citet{fei81}, respectively.}\label{rxknots}
\end{center}
\end{table}

\clearpage

\begin{table}
{\small
\begin{center}
\begin{tabular}{|l|c|c|c|c|c|c|c|c|}\tableline
Knot & ID  & Dist ($'$) & Angle & RA (J2000) & DEC & Counts & Net Rate (cts s$^{-1}\times$10$^3$) & L$_X$ \\ \tableline
 1 & NX1        & 0.09 & 48.58 & 13:25:27.88 & -43:01:06.4 &  198 &  1.44$\pm$0.24 & 1.51E+38 \\ \tableline
 2 & AX1        & 0.25 & 52.94 & 13:25:28.65 & -43:01:00.1 & 1303 & 17.66$\pm$0.50 & 1.84E+39 \\ \tableline
 3 & AX2        & 0.30 & 50.49 & 13:25:28.84 & -43:00:56.9 &  234 &  2.93$\pm$0.22 & 3.07E+38 \\ \tableline
 4 & AX3        & 0.42 & 57.70 & 13:25:29.51 & -43:00:55.9 &  201 &  2.71$\pm$0.20 & 2.84E+38 \\ \tableline
 5 & AX4        & 0.47 & 55.05 & 13:25:29.70 & -43:00:53.8 &  166 &  2.22$\pm$0.18 & 2.23E+38 \\ \tableline
 6 & AX5$^\dag$ & 0.59 & 52.89 & 13:25:30.09 & -43:00:48.2 &   94 &  1.27$\pm$0.14 & 1.33E+38 \\ \tableline
 7 & AX6        & 0.71 & 52.93 & 13:25:30.76 & -43:00:43.6 &  209 &  2.86$\pm$0.20 & 2.99E+38 \\ \tableline
 8 & BX1        & 0.94 & 49.97 & 13:25:31.50 & -43:00:33.4 &  155 &  2.10$\pm$0.17 & 7.40E+37 \\ \tableline
 9 & BX2        & 0.96 & 57.19 & 13:25:32.04 & -43:00:38.0 &  830 & 11.39$\pm$0.40 & 4.01E+38 \\ \tableline
10 & BX3        & 1.03 & 51.45 & 13:25:31.85 & -43:00:29.2 &   71 &  0.90$\pm$0.12 & 3.16E+37 \\ \tableline
11 & BX4        & 1.03 & 60.28 & 13:25:32.52 & -43:00:38.0 &  242 &  3.26$\pm$0.22 & 1.15E+38 \\ \tableline
12 & BX5        & 1.06 & 56.00 & 13:25:32.43 & -43:00:33.4 &  193 &  2.56$\pm$0.20 & 9.01E+37 \\ \tableline
13 & CX1$^\dag$ & 1.14 & 56.71 & 13:25:32.88 & -43:00:31.0 &  131 &  1.67$\pm$0.16 & 5.89E+37 \\ \tableline
14 & CX2        & 1.23 & 47.43 & 13:25:32.56 & -43:00:19.4 &  174 &  2.27$\pm$0.19 & 7.98E+37 \\ \tableline
15 & CX3        & 1.29 & 58.28 & 13:25:33.74 & -43:00:28.5 &   57 &  0.72$\pm$0.11 & 2.53E+37 \\ \tableline
16 & CX4        & 1.35 & 53.16 & 13:25:33.52 & -43:00:20.1 &   93 &  1.23$\pm$0.14 & 4.33E+37 \\ \tableline
17 & EX1$^\dag$ & 1.54 & 50.60 & 13:25:34.13 & -43:00:10.2 &   40 &  0.47$\pm$0.09 & 1.66E+37 \\ \tableline
18 & EX2        & 1.63 & 53.79 & 13:25:34.80 & -43:00:10.9 &   39 &  0.50$\pm$0.09 & 1.75E+37 \\ \tableline
19 & EX3        & 1.73 & 51.15 & 13:25:34.86 & -43:00:04.3 &   49 &  0.57$\pm$0.10 & 2.00E+37 \\ \tableline
20 & FX1        & 1.85 & 58.82 & 13:25:36.27 & -43:00:11.6 &   80 &  1.05$\pm$0.13 & 3.70E+37 \\ \tableline
21 & FX2        & 1.87 & 54.26 & 13:25:35.95 & -43:00:03.2 &   86 &  1.15$\pm$0.13 & 4.04E+37 \\ \tableline
22 & FX3        & 1.96 & 58.01 & 13:25:36.72 & -43:00:06.7 &   77 &  1.05$\pm$0.12 & 3.70E+37 \\ \tableline
23 & FX4        & 2.03 & 57.56 & 13:25:36.98 & -42:00:03.9 &   44 &  0.55$\pm$0.10 & 1.95E+37 \\ \tableline
24 & FX5        & 2.15 & 58.59 & 13:25:37.65 & -43:00:01.8 &   37 &  0.50$\pm$0.09 & 1.75E+37 \\ \tableline
25 & FX6        & 2.24 & 53.32 & 13:25:37.46 & -42:59:49.1 &   69 &  0.91$\pm$0.12 & 3.21E+37 \\ \tableline
26 & FX7        & 2.51 & 51.09 & 13:25:38.32 & -42:59:34.7 &   58 &  0.77$\pm$0.11 & 2.73E+37 \\ \tableline
27 & GX1        & 3.02 & 55.17 & 13:25:41.17 & -42:59:25.6 &   57 &  0.77$\pm$0.11 & 2.73E+37 \\ \tableline
28 & GX2        & 3.13 & 53.75 & 13:25:41.43 & -42:59:18.2 &   65 &  0.87$\pm$0.11 & 3.07E+37 \\ \tableline
29 & GX3        & 3.26 & 55.05 & 13:25:42.23 & -42:59:17.2 &   97 &  1.32$\pm$0.14 & 4.67E+37 \\ \tableline
30 & GX4        & 3.40 & 54.53 & 13:25:42.78 & -42:59:10.8 &   65 &  0.90$\pm$0.11 & 3.16E+37 \\ \tableline
31 & GX5        & 3.75 & 53.27 & 13:25:44.12 & -42:58:55.0 &   14 &  0.19$\pm$0.05 & 6.81E+36 \\ \tableline
\end{tabular}
\caption{Summary of knots and enhancements detected in the Centaurus A jet.  The distance
is the angular distance from the nucleus in arcminutes, the angle
is the position angle from the nucleus to the knots measured east from
north, the counts are the source counts in a $3.5''$ (inner jet - knots NX1 and AX1 through AX6)
or $4''$ radius circle around each feature in the 0.4-2.5 keV band (not
background subtracted),
the rate is the background-subtracted rate (cts/s) and statistical uncertainty,
and the luminosity is given in the 0.1-10 keV band (unabsorbed) assuming a power-law spectrum
of photon index 2.5 and $N_H$=7.2$\times$10$^{21}$cm$^{-2}$ (inner jet - knots
AX1 through AX6) or photon index 2.3 and $N_H$=1.7$\times$10$^{21}$cm$^{-2}$
(all other knots and enhancements) computed using HEASARC PIMMS.
The background was estimated from a region extending radially from
the nucleus adjacent to the jet.
The enhancements labeled with $\dag$ are those which are consistent
with point sources as defined by the
encircled energy test described in the text, and are therefore possibly
XRBs unrelated to the jet.}\label{knottab}
\end{center}
}
\end{table}

\clearpage

\begin{table}
\begin{center}
\begin{tabular}{|c|c|c|c|}\tableline
Projection  & Inner radius (arcsec) & Outer radius (arcsec) & Approx. location \\ \tableline
 1          &  20.7 &  48.2 & Knots AX1-AX4 \\ \tableline
 2          &  48.2 &  76.8 & Knot B        \\ \tableline
 3          &  76.8 & 124.5 & Knots C and E \\ \tableline
 4          & 124.5 & 214.9 & Knots F and G \\ \tableline
\end{tabular}
\caption{Summary of projected regions of the Cen A jet.  We have determined
the width of the jet in projection (rotated through position angle 55$^\circ$)
in two energy bands (0.4-1.5 keV and 1.5-5 keV) for each of these regions.
The projections of the first two regions are shown in Figure~\ref{proj1}.  We find that there is
no difference in the width of the jet in the two energy bands in these four regions.}\label{projregs}
\end{center}
\end{table}

\clearpage

\begin{table}
\begin{center}
\begin{tabular}{|l|c|c|c|c|c|}\tableline
Feature       & Box                & Rate (cts/s)                 & Phot. Index & \nh (PL)       & Luminosity \\ \tableline
Knots AX1/AX2   & $0.18' \times 0.11'$ & 3.7$\pm$0.2$\times$10$^{-2}$ & 2.5$\pm$0.05 & 7.2$\pm$0.3E21 & 3.31E39 ergs s$^{-1}$ \\ \tableline
Knots AX3-AX6   & $0.48' \times 0.22'$ & 2.6$\pm$0.3$\times$10$^{-2}$ & 2.2$\pm$0.1 & 6.7$\pm$0.8E21 & 1.45E39 ergs s$^{-1}$ \\ \tableline
Knot B          & $0.38' \times 0.27'$ & 5.3$\pm$0.3$\times$10$^{-2}$ & 2.0$\pm$0.05 & 2.0$\pm$0.3E21 & 1.47E39 ergs s$^{-1}$ \\ \tableline
Beyond knot B   & $1.93' \times 0.43'$ & 4.1$\pm$0.2$\times$10$^{-2}$ & 2.3$\pm$0.1 & 1.7$\pm$0.5E21 & 1.27E39 ergs s$^{-1}$ \\ \tableline
\end{tabular}
\caption{Summary of the best-fit (0.4 - 5 keV) parameters for four regions of
Centaurus A jet using an absorbed power-law model.  The first
column identifies the feature, the second is the
size of the region around the feature used for spectral
analysis, the third column is the background-subtracted count
rate from the feature and the statistical (one $\sigma$) uncertainty,
the fourth column is the best-fit photon index and uncertainty,
the fifth is the best-fit absorbing column (cm$^{-2}$), and
the sixth is the intrinsic (i.e. unabsorbed) luminosity in 
the 0.1-10 keV bandpass.  The Galactic column density in the direction of
Cen A is 7$\times 10^{20}$ cm$^{-2}$.
All uncertainties on best-fit parameters are the 1$\sigma$ confidence intervals for two 
interesting parameters.}\label{sfit}
\end{center}
\end{table}

\clearpage

\begin{table}
\begin{center}
\begin{tabular}{|l|c|c|c|} \tableline
Knot  &  Radio (mJy)  & X-ray (nJy) & $\alpha _{rx}$ \\ \tableline
\multicolumn{4}{|c|}{X-ray Peak Centered} \\ \tableline
NX1   &  23    &     5.8   &    0.88 \\ \tableline
AX1   & 333    &    110    &    0.86 \\ \tableline
AX2   & 247    &    14     &    0.97 \\ \tableline
AX3   & 450    &    28     &    0.97 \\ \tableline
AX4   & 199    &    14     &    0.96 \\ \tableline
AX6   &  46    &    23     &    0.84 \\ \tableline
BX2   &  58    &    66     &    0.80 \\ \tableline
BX5   & 497    &    50     &    0.94 \\ \tableline
\multicolumn{4}{|c|}{Radio Peak Centered} \\ \tableline
A1r   & 324    &    106    &    0.87 \\ \tableline
A2r   & 519    &    22     &    0.99 \\ \tableline
A3/4r & 503    &    15     &    1.01 \\ \tableline
\end{tabular}
\caption{Summary of radio to X-ray spectral indices for knots
of the inner jet region.  The first column is the knot label,
the second is the radio (8.4 GHz) flux density, the third
is the X-ray (1 keV) flux density, and the fourth is the
radio to X-ray energy spectral index.  Each of these flux densities
and spectral indices was computed from a box region $4''$ on
a side.}\label{fdens}
\end{center}
\end{table}

\end{document}